\documentclass[smallextended]{svjour3}       % onecolumn (second format)
\smartqed  % flush right qed marks, e.g. at end of proof

\RequirePackage{fix-cm}

\usepackage{graphicx}
\usepackage{listings}
\usepackage{pifont}
\usepackage{amsmath}
\usepackage{pdflscape} 
\usepackage{tikz}
\usepackage{dirtytalk}
\usepackage{makecell}
\usepackage[title]{appendix}
\usepackage{pythonhighlight}
\usepackage{caption}
\newenvironment{code}{\captionsetup{type=listing}}{}
\definecolor{LightGray}{gray}{0.9}
\usepackage[autostyle]{csquotes}
\usepackage{longtable}
\usepackage{booktabs}
\usepackage{tablefootnote}
\usepackage{multirow}
\usepackage{adjustbox}
\usepackage{enumitem}
\usepackage{siunitx}
\usepackage{fancyvrb}
\usepackage{soul}
\usepackage{textcomp}
\usepackage{tabularx}
\newlength\someheight
\usepackage{pgfplots}
\usepackage{pgfplotstable}
\pgfplotsset{compat=1.14}
\definecolor{storeClusterComponent}{HTML}{808080}
\definecolor{dbscan}{HTML}{BEBEBE}
\definecolor{constructCluster}{HTML}{DCDCDC}
\usepackage{multirow}
\usepackage{tikz} 
\usepackage{subcaption}
\usetikzlibrary{shapes,calc,shadows,arrows}
\usepackage{pgf-pie}
\usepackage{etoolbox}
\newtoggle{showpct}
\usepackage{xcolor,colortbl}

\definecolor{codegreen}{rgb}{0,0.6,0}
\definecolor{codegray}{rgb}{0.5,0.5,0.5}
\definecolor{codepurple}{rgb}{0.58,0,0.82}
\definecolor{backcolour}{rgb}{0.95,0.95,0.92}

\lstdefinestyle{mystyle}{
    backgroundcolor=\color{backcolour},   
    commentstyle=\color{codegreen},
    keywordstyle=\color{magenta},
    numberstyle=\tiny\color{codegray},
    stringstyle=\color{codepurple},
    basicstyle=\ttfamily\footnotesize,
    breakatwhitespace=false,         
    breaklines=true,                 
    captionpos=b,                    
    keepspaces=true,                 
    numbers=left,                    
    numbersep=5pt,                  
    showspaces=false,                
    showstringspaces=false,
    showtabs=false,                  
    tabsize=2
}
\usepackage[most]{tcolorbox}

\newtcbtheorem{Summary}{\bfseries Summary}{enhanced,drop shadow={black!50!white},
  coltitle=black,
  top=0.3in,
  attach boxed title to top left=
  {xshift=1.5em,yshift=-\tcboxedtitleheight/2},
  boxed title style={size=small,colback=pink}
}{summary}

\newtcolorbox[auto counter]{summary}[1][]{title={\bfseries Summary~\thetcbcounter},enhanced,drop shadow={black!50!white},
  coltitle=black,
  top=0.3in,
  attach boxed title to top left=
  {xshift=1.5em,yshift=-\tcboxedtitleheight/2},
  boxed title style={size=small,colback=pink},#1}

\usepackage{natbib}

\usepackage{etoolbox}
\usepackage[hyphens]{url}
\usepackage{hyperref}
\PassOptionsToPackage{hyphens}{url}
\hypersetup{colorlinks,
    citecolor=blue, 
    urlcolor=black,
    linkcolor=blue}

\newcommand*\circled[1]{\tikz[baseline=(char.base)]{
            \node[shape=circle,draw,inner sep=1pt] (char) {#1};}}
\makeatletter
\patchcmd{\pgfpie@slice}%
{\scalefont{#3}\beforenumber#3\afternumber}%
{\iftoggle{showpct}{\scalefont{#3}\beforenumber#3\afternumber}{}}%
{}{}

\patchcmd{\NAT@citex}
  {\@citea\NAT@hyper@{%
     \NAT@nmfmt{\NAT@nm}%
     \hyper@natlinkbreak{\NAT@aysep\NAT@spacechar}{\@citeb\@extra@b@citeb}%
     \NAT@date}}
  {\@citea\NAT@nmfmt{\NAT@nm}%
   \NAT@aysep\NAT@spacechar\NAT@hyper@{\NAT@date}}{}{}

\patchcmd{\NAT@citex}
  {\@citea\NAT@hyper@{%
     \NAT@nmfmt{\NAT@nm}%
     \hyper@natlinkbreak{\NAT@spacechar\NAT@@open\if*#1*\else#1\NAT@spacechar\fi}%
       {\@citeb\@extra@b@citeb}%
     \NAT@date}}
  {\@citea\NAT@nmfmt{\NAT@nm}%
   \NAT@spacechar\NAT@@open\if*#1*\else#1\NAT@spacechar\fi\NAT@hyper@{\NAT@date}}
  {}{}

\def\srcfile[#1,#2]#3{
    \node [draw, fill=white, minimum height=1.7cm, minimum width=1.35cm, rounded corners, double copy shadow={shadow xshift=0.1cm, shadow yshift=0.1cm}] (#1) at #2 {};
    \node[align=center, font=\sffamily\fontsize{5}{1.5}\selectfont] at #2 {#3};
}

\makeatother

\definecolor{dkgreen}{rgb}{0,0.6,0}
\definecolor{gray}{rgb}{0.5,0.5,0.5}
\definecolor{mauve}{rgb}{0.58,0,0.82}
\definecolor{dgreen}{rgb}{0.0, 0.5, 0.0}

\usepackage{mdframed}
\newcounter{finding}
\newmdenv[%
    linewidth=0.6pt,
    linecolor=black,
    outerlinewidth=0pt,
    skipabove=0pt,
    skipbelow=0pt,
    settings={\global\refstepcounter{finding}},
]{myfinding}

\newcommand{\finding}[1]{
    \vspace{0.5em}
    \begin{myfinding}
	\textbf{\textit{Finding~\arabic{finding}}}: #1
    \end{myfinding}
    \vspace{0.5em}
}

\makeatletter
\newcommand\notsotiny{\@setfontsize\notsotiny\@vipt\@viipt}
\makeatother

\usepackage[draft=true]{minted}
\newminted{diff}{
    fontsize=\scriptsize,
	autogobble=true,
}
\newmintinline{cpp}{}
\newminted{python}{}
\newmintinline{python}{}
\newminted{java}{
	fontsize=\small,
	autogobble=true,
	breaklines=true,
	breakafter=.,
	breakautoindent,
    tabsize=2,
    breakbytoken=true,
    resetmargins=true}
\newmintinline{java}{
    fontsize=\small}
\setminted{%
    style=trac,
    autogobble=true,
    breaklines=true,
    breakafter=.,
    breakautoindent,
    escapeinside=||}

\DeclareCaptionSubType{listing}
\newcommand{\ji}[1]{\javainline{#1}}

\usepackage{amsthm}

\theoremstyle{definition}

\usepackage{threeparttable}

\definecolor{dodgerblue}{RGB}{30,144,255}
\definecolor{orange}{RGB}{255, 120, 8}

\newcommand{\lstbg}[3][0pt]{{\fboxsep#1\colorbox{#2}{\strut #3}}}
\lstdefinelanguage{diff}{
  basicstyle=\ttfamily\small,
  morecomment=[f][\lstbg{red!20}]-,
  morecomment=[f][\lstbg{green!20}]+,
  morecomment=[f][\textit]{@@},
}

\newcolumntype{C}[1]{>{\centering\arraybackslash}m{#1}}

\tcbset{commonstyle/.style={boxrule=0pt,sharp corners,enhanced jigsaw,nobeforeafter,boxsep=0pt,left=\fboxsep,right=\fboxsep}}
\newtcolorbox{mycolorbox}[1][]{commonstyle,#1}

\usepackage{svg}

\usepackage{etoolbox}
\newcommand*{\affaddr}[1]{#1} % No op here. Customize it for different styles.
\newcommand*{\affmark}[1][*]{\textsuperscript{#1}}

\usepackage{scalerel}
\usepackage{tikz}
\usetikzlibrary{svg.path}

\definecolor{orcidlogocol}{HTML}{A6CE39}
\tikzset{
  orcidlogo/.pic={
    \fill[orcidlogocol] svg{M256,128c0,70.7-57.3,128-128,128C57.3,256,0,198.7,0,128C0,57.3,57.3,0,128,0C198.7,0,256,57.3,256,128z};
    \fill[white] svg{M86.3,186.2H70.9V79.1h15.4v48.4V186.2z}
                 svg{M108.9,79.1h41.6c39.6,0,57,28.3,57,53.6c0,27.5-21.5,53.6-56.8,53.6h-41.8V79.1z M124.3,172.4h24.5c34.9,0,42.9-26.5,42.9-39.7c0-21.5-13.7-39.7-43.7-39.7h-23.7V172.4z}
                 svg{M88.7,56.8c0,5.5-4.5,10.1-10.1,10.1c-5.6,0-10.1-4.6-10.1-10.1c0-5.6,4.5-10.1,10.1-10.1C84.2,46.7,88.7,51.3,88.7,56.8z};
  }
}

\newcommand\orcidicon[1]{\href{https://orcid.org/#1}{\mbox{\scalerel*{
\begin{tikzpicture}[yscale=-1,transform shape]
\pic{orcidlogo};
\end{tikzpicture}
}{|}}}}

\usepackage{hyperref}

\journalname{Empirical Software Engineering}
\begin{document}

\title{Studying Logging Practice in Machine Learning-based Applications}

\author{Patrick Loic Foalem\and Foutse Khomh \and Heng Li
}

\authorrunning{Patrick Loic Foalem\and Foutse Khomh \and Heng Li}

%\authorrunning{Short form of author list} % if too long for running head

\institute{ \affmark[*]Corresponding author. \\
\\
           Patrick Loic Foalem, Foutse Khomh, Heng Li \at
              \affaddr{Department of Computer Engineering and Software Engineering \\ Polytechnique Montreal \\
              Montreal, QC, Canada} \\
              \email{\{patrick-loic.foalem, foutse.khomh, heng.li\}@polymtl.ca}           %  \\
%             \emph{Present address:} of F. Author  %  if needed
}

\date{Received: date / Accepted: date}
% The correct dates will be entered by the editor

\maketitle

\begin{abstract}
% Insert your abstract here. Include keywords, PACS and mathematical
% subject classification numbers as needed.

Logging is a common practice in traditional software development. Several research works have been done to investigate the different characteristics of logging practices in traditional software systems (e.g., Android applications, JAVA applications, C/C++ applications). Nowadays, we are witnessing more and more development of Machine Learning-based applications (ML-based applications). Today, there are many popular libraries that facilitate and contribute to the development of such applications, among which we can mention: Pytorch, Tensorflow, Theano, MXNet, Scikit-Learn, Caffe, and Keras. Despite the popularity of ML, we don't have a clear understanding of logging practices in ML applications. In this paper, we aim to fill this knowledge gap and help ML practitioners understand the characteristics of logging in ML-based applications. %In this paper we are trying to fill this knowledge gap.
% \Foutse{methodology is mixed up with results which is mix up with implication....please use a structured abstract, by dividing into :\\ - Context; - Objective; - Method; - Results; 
% - Conclusions....it will make it more coherent!}
In particular, we conduct an empirical study on 110 open-source ML-based applications. Through a quantitative analysis, we find that logging practice in ML-based applications is less pervasive than in traditional applications including Android, JAVA, and C/C++ applications. Furthermore, the majority of logging statements in ML-based applications are in \textit{info} and \textit{warn} levels, compared to traditional applications where \textit{info} is the majority of logging statement in C/C++ application and \textit{debug}, \textit{error} levels constitute the majority of logging statement in Android application. We also perform a quantitative and qualitative analysis of a random sample of logging statements to understand where ML developers put most of logging statements and examine why and how they are using logging. These analyses led to the following observations: (i) ML developers put most of the logging statements in \textit{model training}, and in \textit{non-ML components}. (ii) Data and model management appear to be the  main reason behind the introduction of logging statements in ML-based applications. Indeed, ML developers use logging statements to keep track of the different stages of data processing as well as record relevant statistical information. In addition, they use logging statements to track all the experiments performed, to find the best model, and to record all the performances achieved by the model on validation or test data. Through this work, we hope to shed light on logging practices in ML-based applications. %, our results make ML developers aware not only of the existence of different logging libraries but also of their use in ML applications. 
Our results also highlight the need for more ML-specific logging libraries
% \Foutse{did we found that adequate libraries for logging is missing? can you elaborate?} 
to record and automate all experiments performed in the pipeline of ML-based applications. %, since the majority of logging statement used in ML-based applications currently come from non ML logging libraries.
% \heng{explain the two reasons}\Foutse{yes please! you can also add a line about the potential impact of your work....how can people use your results...etc} 
% %with some example of logging statement to show what are they logging.
% \heng{Add key take-home messages: how the results benefit practitioners/researchers/logging library providers}

% \PACS{PACS code1 \and PACS code2 \and more}
% \subclass{LOCC code1 \and LOCC code2 \and more}
\end{abstract}
\section{Introduction}
\label{sec:section1}
Inserting logging statements into software programs is part of good programming practice (\citet{loggingpractice}). %They 
Logging statements allow to collect information about the behavior of a program during its execution. This information is often used for various tasks in software maintenance and operations, such as diagnosing failures, reporting error, detecting anomalies, and automatically generating logging text using log data \citep{yang2016log, nagaraj2012structured, xu2009detecting, 9825813}. Many research studies have been conducted to characterize logging practices in traditional software development, for example, \citet{yuan2012characterizing} studied logging practices in open-source applications written in C/C++ , \citet{zeng2019studying} made similar study for android applications, and \citet{chen2017characterizing} studied logging practices in JAVA-based applications. However, to the best of our knowledge, no previous research work examined %we could not find any research work focusing on 
logging practices in ML-based applications. This paper aims to fill this gap in the literature. Such a study could be useful for ML practitioners to understand logging practices in ML applications, identify the kind of information that is being logged, and the logging libraries used.
\noindent
Logs are generated by logging statements. Listing \autoref{code:c-code} shows some typical examples of logging statements used in ML applications. The logging statements from lines 1, 3, and 9 use general logging libraries of the Python programming language, while the logging statements on lines 5, 7, 11, 13, 15 use ML-specific logging libraries. %logging statements for ML software (e.g., line 5, 7, 11, 13, 15). 
General logging libraries are built from Python module \ji{logging}. Their logging statement always have log level (\ji{INFO, Warn, ERROR, DEBUG, CRITICAL}) and in general, these libraries are older than ML-specific logging libraries. They are designed primarily for traditional software systems, %. These ldevelopers and have logging levels, 
whereas ML-specific logging libraries are designed for ML applications and contain information specific to machine learning and offer a straightforward approach for logging performance, hyperparameters, and statistical information about the data. Some ML logging libraries even offer the advantage of visualizing the logged elements in a graph, allowing users to browse through the learning process of the model.  
% do not necessarily have log level \Foutse{instead of saying that they dont have something instead say what they have!!!} (\ji{INFO, Warn, ERROR, DEBUG, CRITICAL}). 
%\heng{Explain the difference between the general logging and ML-specific logging}

\begin{center}
% \Foutse{add a caption and refer to this listing in the text!} \Patrick{Correct??}
\begin{code}
\begin{minted}[frame=lines,
framesep=2mm,
baselinestretch=1,
bgcolor=LightGray,
fontsize=\footnotesize,
linenos]{python}
logging.info("*** Reading from input files ***")

tf.logging.info("[*] num_epochs: %d" % _num_epochs)

dllogger.log(step=(epoch, steps_per_epoch, batch_idx), data=dllogger_data)

caffe.log('Using devices %s' % str(gpus))

rospy.logwarn("Updating Steering PID %s, %s, %s", config["Steer_P"], config["Steer_I"], config["Steer_D"])

self.hparams.train_logger.log_stats({"Epoch": epoch, "lr": old_lr}, train_stats={"loss": self.train_loss}, valid_stats=stats, )

mlflow.log_metrics(metrics, step=step)

wandb.log(log_data, step=learner.num_samples_collected + args.start_env_steps)
\end{minted}
\captionof{listing}{Logging statement examples extracted from ML-based open-source application.}
\label{code:c-code}
\end{code}
\end{center}

To understand logging practices in ML-based applications, in this paper, we conduct an empirical study of 110 ML-based software projects from GitHub\footnote{\url{https://github.com/}}; examining the prevalence of logging statements, the rationale for %reason 
their use, and the characteristics of the ML components in which logging occur. % are more prone to logging statements.
%This study is conducted on 110 ML-based software projects with source code hosted on GitHub\footnote{\url{https://github.com/}}. These 
% \Foutse{this is too much details for an introduction, please move it to data collection section!!!}The studied projects use at least one of the following libraries : TensorFlow\footnote{\url{https://www.tensorflow.org/}}, Theano\footnote{\url{https://theano-pymc.readthedocs.io/en/latest/}}, MXNet\footnote{\url{https://mxnet.apache.org/versions/1.9.1/}}, Scikit-Learn\footnote{\url{https://scikit-learn.org/stable/}}, Pytorch\footnote{\url{https://pytorch.org/}}, Keras\footnote{\url{https://keras.io/}},  Caffe\footnote{\url{https://caffe.berkeleyvision.org/}} which are the most popular libraries in the field of machine learning \citep{islam2019comprehensive,wang2020deep}. 
%The summary of our research questions and findings are as follows:
Our research is organized along three following research questions (RQs): % as described below.

\begin{enumerate}[start=1,label={\bfseries RQ\arabic*:}, leftmargin = 1em]
  \item \textit{What are the characteristics of ML-based software logging practices?}  
%   \Foutse{don't put boxes in the introduction!}
  
%   \begin{tcolorbox}[colback=gray!20,colframe=blue!40!black]
  To answer this research question, we quantify the logging statements present in our studied ML-based systems and compare the result with previous findings on traditional systems (i.e., JAVA, C/C++, Android). We observe a difference between the density of logs contained in %By measuring log density, we observe that logging practices in 
  ML-based applications and the density reported for traditional %re different from those 
  Android, JAVA, and C/C++ applications. In particular, logging in ML-based applications is less pervasive than in JAVA, Android, and C/C++ applications. Moreover, the distribution of logging levels in ML-based applications are different from those in traditional applications.
% \end{tcolorbox}
  \item \textit{Which phases of the ML pipeline are more prone to logging?}
  
% \begin{tcolorbox}[colback=gray!20,colframe=blue!40!black]
  The ML pipeline consists of the following steps: data collection, data processing, model training, model evaluation, and model deployment. We manually analyzed $\sim$2K logging statements that corresponds to more than 95\% of confidence level with a 5\% confidence interval \citep{yamane1967statistics}. We mapped each logging statement to one stage of the ML pipeline and examined their content. %mapped assigned each logging statement to an ML step. Our results show that, 
  We observed that the majority of logging statements occurred %found 
  in the model training step and contains information about the models' hyperparameters and performances.
  
%   .\Foutse{and what are they about? what is in these logs?}
%   \Foutse{you didnt talk that you check components!}.
% \end{tcolorbox}
  \item \textit{Why do %are the reason 
  developers use logging in ML-based application?}
  
% \begin{tcolorbox}[colback=gray!20,colframe=blue!40!black]
 %Although we confirm that logging is widely used in ML-based applications, from our sample we tried to 
 To understand the rationale behind development decisions for using logging statements, we examined the code of file containing 380 logging statements from our sample, and assigned a reason behind its use. 
%  we identify what kind of information is logged and the reason \Foutse{how do you infer these reasons?} behind its use. Then categorize them according to the %in two main reasons for using logging in ML-based applications. 
 Results show that ML practitioners use logging statements for two mains reasons : data and model management.
%  \Foutse{so they don't do any of the other logging operations of traditional systems? yes of course they will log some state event for debugging purpose but the main reason behind that is to manage data processing or model training phase i speak about that further in the paper} %reasons why developers of ML systems use logging statements. 
%  \Foutse{please revise the following sentence! just briefly report the key finding for this RQ!}We have identified two mains reasons for which logging statements are used, providing examples of their use. Its reasons are : data and model management.
% \end{tcolorbox}  
\end{enumerate}

These results show a distinct logging practice in ML-based applications from widely-studied traditional applications and call for more research work to improve our understanding of logging practices in ML-based applications. The high proportion of logging statements related to general logging libraries (used for code and model management) in ML code may be a sign that ML practitioners are not aware of ML-specific logging libraries, such as \ji{mlflow, wandb}. % such as \Foutse{can you name some ML-specific libraries?} \Foutse{and do they offer better features? please explain..if not, one can say, why use them then!}. %and the main purpose of logging statements in ML applications. 
As an example, in Listing \autoref{code:c-cod}, ML practitioners continued to use the general logging libraries to log the performance of the models during its training phase i.e., line 1, instead of using ML-specific logging libraries designed to log this type of information despite the advantage offered by these ML-specific libraries; e.g., being able to log many ML metrics at once, tracking all the experiments performed during the learning phase and summarizing them through visualizations, in order to track performance improvements. i.e., line 3, 5, 6 and 8.
% \Foutse{more efficiently? (what are the benefits of those domain specific logging libraries?)} \Foutse{depending on how the general purpose logging libraries are used...it could also signal that current ML logging libraries are missing some important features! and people try to get them from general purpose ones even if they don't fit perfectly...i think you should examine a bit to see if this argument holds}.

\begin{center}
% \Foutse{add a caption and refer to this listing in the text!} \Patrick{Correct??}
\begin{code}
\begin{minted}[frame=lines,
framesep=2mm,
baselinestretch=1,
bgcolor=LightGray,
fontsize=\footnotesize,
linenos]{python}
logger.info('Epoch {} Loss {:.4f} Accuracy {:.4f}'.format( epoch,
loss_meter.avg, accuracy))
dllogger.log(step=(epoch, steps per epoch, batch idx),
data=dllogger data, verbosity=0)
wandb.log({"RMSE_training": training_RMSE, "epoch": epoch,})
self.hparams.train_logger.log_stats({"Epoch": epoch, "lr": old_lr},
train_stats={"loss": self.train_loss}, valid_stats=stats, )
mlflow.log_metrics(metrics, step=step)
\end{minted}
\captionof{listing}{Logging statement examples use for logging ML performance.}
\label{code:c-cod}
\end{code}
\end{center}
% These results show that \Foutse{please formulate a generic lesson that can be derived from your results!!! and briefly explain how practitioners and researchers can use it!}
%The implications of our results are showing distinct logging practices in ML-based applications from the widely-studied traditional application. However, due to the nondeterministic nature of AI algorithms, it is essential for ML practitioners and future research to better address the challenges of logging practices in ML applications. Also logging tools to support the development of AI component are required.

\paragraph{\textbf{\textup{Paper organization.}}}
The remainder of this paper is organized as follows. Section \ref{sec:section2} describes the design of our study. %introduces our case study setup. 
Section \ref{sec:section3} presents the motivation, results of each research question. Section \ref{sec:threatsToValidity} discusses threats to the validity of our findings. Section \ref{sec:relatedwork} discusses the related literature and Section \ref{sec:conclusion} concludes the paper.

\begin{figure}[htp!]
\includegraphics[width=\textwidth]{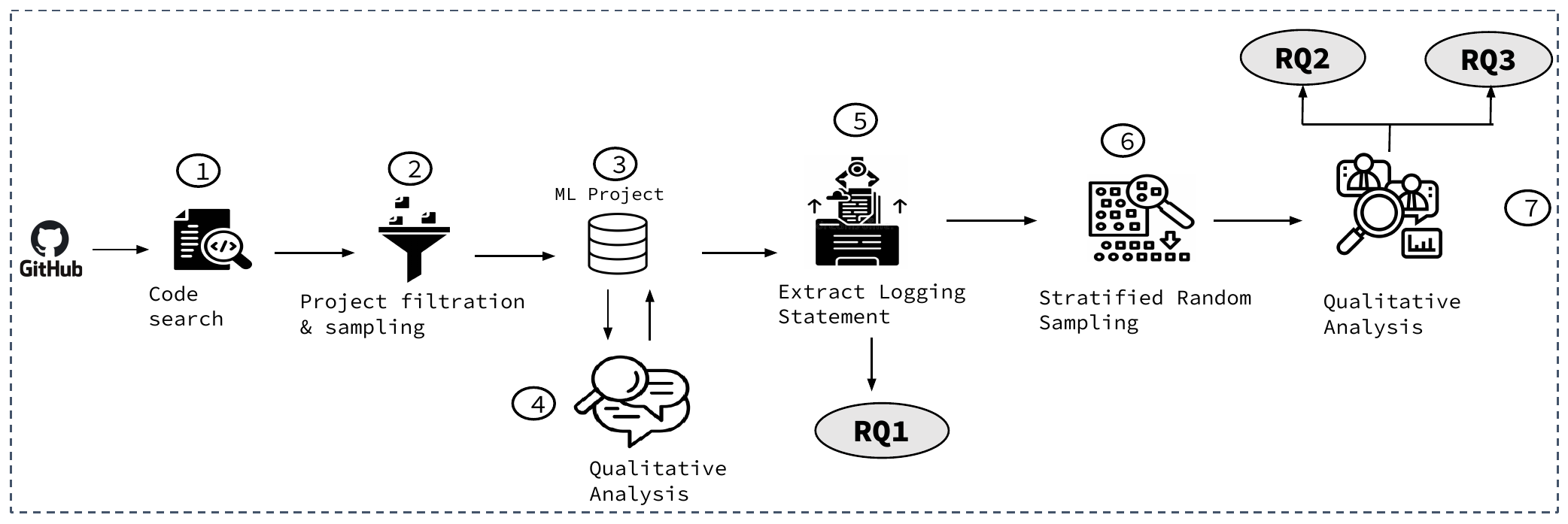}
\centering
\caption{Overview of the research workflow.} 
\label{fig:researchprocess} 
\end{figure}
\section{Study Design}
\label{sec:section2}
This section presents the design of our study, which aims to understand logging practices in ML-based applications.
\subsection{Data collection}
\label{sec:datacollection}
%In this paper, we set out to empirically examine logging practices in ML-based applications. In this section, we describe our projects selection process, and present the methodology followed to extract our data sets. %of data extraction. 
\autoref{fig:researchprocess} provides an overview of our data extraction and analysis approach. We describe each step below. 
%In  and dagives an overview of our approach. We divide our methodology in seven main steps. Each step is described in the following sub-sections.
%\footnote{Scripts and data files used in our research are available online and can be found here: \url{https://github.com/senseconcordia/TestLoggingPractice}}.

% \autoref{fig:researchprocess} illustrates an overview of our research workflow. 
% In each analyzed subject, we examined the logging practice in its latest version and all commit histories. The workflow mainly consists of seven steps.

\begin{enumerate}
    \item[\ding{172}] \textbf{Code search}

To select ML-based software projects for our study, we used the GitHub API. GitHub Search API V3\footnote{ \url{https://docs.github.com/en/rest}} allows us to collect projects by doing code search. We searched for projects based on the \texttt{import} statements. To obtain ML-based projects, we follow the same approach as  \cite{OpenjaMKCL22}, which consists in searching for machine learning libraries among the import statements of GitHub projects. Specifically, we search for the  import statements corresponding to the following ML libraries:  %such as \Foutse{are there more? why do you say such as? if there are more, you should say where to find the full list!}
TensorFlow, Theano, MXNet, Scikit-Learn, Pytorch, Keras, Caffe. In total, we collect 1,547 ML-based projects. We focus on these libraries because of their popularity.   
    \item[\ding{173}] \textbf{Project filtration \& sampling}.
    
In order to retain a relevant quantity of mature projects for our study, we apply the filtering approach proposed by previous works \citep{OpenjaMKCL22,munaiah2017curating}, which consists in defining thresholds on the following three metrics: number of stars, number of commits, number of authors. To identify adequate threshold values, we plot the distributions of the three metrics.  % for each of the metrics. 
\autoref{fig:y equals x} represents the cumulative frequency curve of all projects that we collected from GitHub by number of commits. For this metric (i.e., number of commits), we define the threshold value to be % of for this metric at 
100 to avoid \say{toy} projects such as tutorial, student's class assignments. Using this threshold we retained 320 projects. We also plotted the cumulative frequency curve for the other two metrics (see the Appendix \autoref{img12} for the corresponding figures) and derived a threshold value of one for the %of at least one for the 
number of stars and two for the number of authors.
We then filtered out all projects in our dataset with less than one star and less than two authors, and retained  a final set of 110 projects. 
% compute \Foutse{what?} and visualised the \Foutse{...please explain!}, as shown on % have performed the plots in 
% \autoref{fig:three graphs}. These figures presents the cumulative \Foutse{what? be precise please...curve doesnt mean anything!} curves of the projects obtained in the previous step. % according to some metrics present in \autoref{tab:table1}. 
% \Foutse{which criteria did you infer from these figures and why?, also cite other papers that followed a similar methodology!}. Hence, we % Based on this visualization, we 
% applied the following selection criteria: (1) a project should have at least one stars, (2) a project should be implemented by at least 2 authors, and (2) a project should have at least one hundred commits.

     \begin{figure}[htp]
         \centering
         \includegraphics[width=\textwidth]{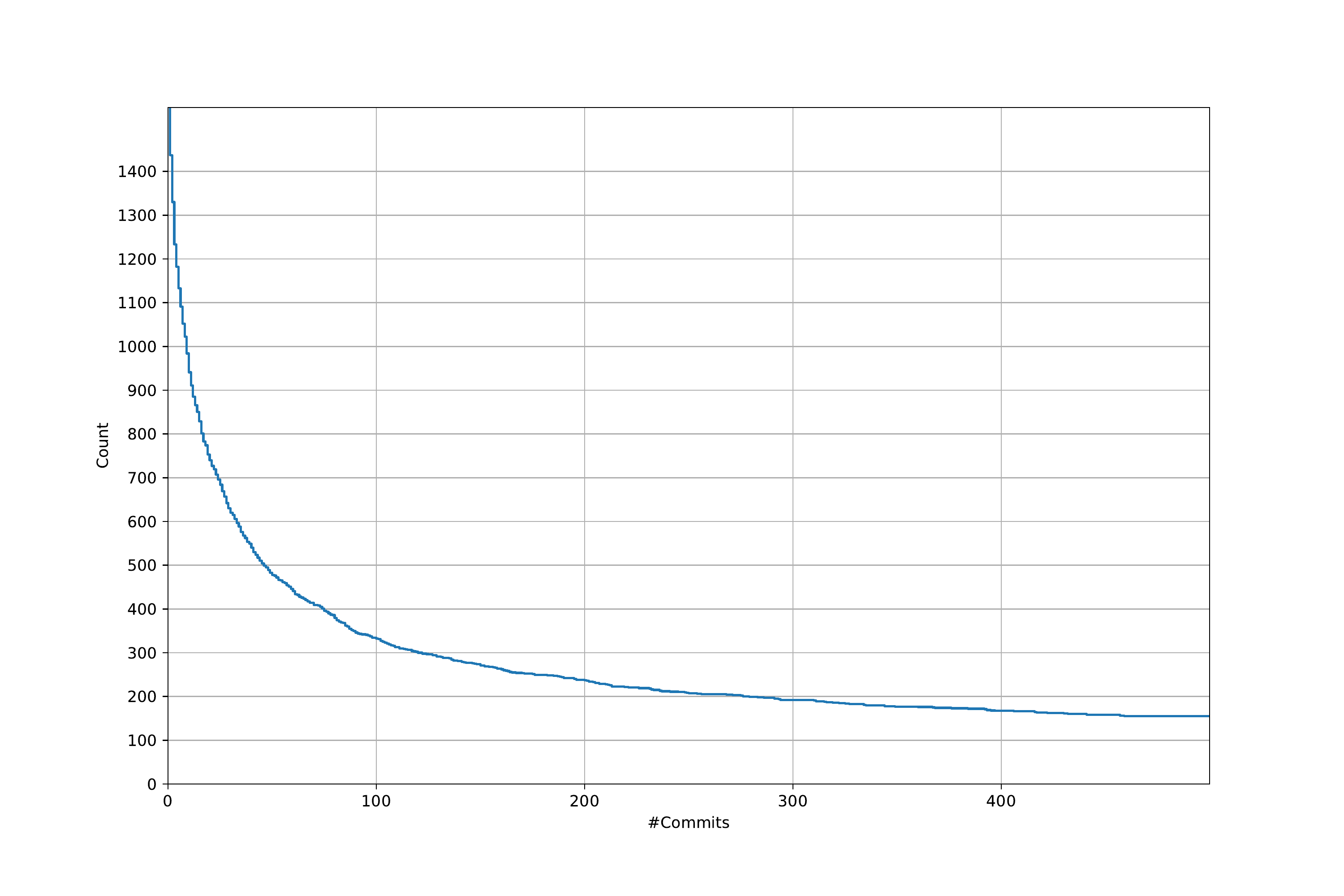}
         \caption{Cumulative frequency curve base on \#Commits}
         \label{fig:y equals x}
     \end{figure}
        % \caption{Cumulative curve of our dataset before filtering on three metric \Foutse{may be just show one figure to explain how you obtained the cut -off point and then explain that you repeated the process for the other metrics and report their corresponding thresholds....this way you can show only one figure here and move the other figures to the appendix }}

    \item[\ding{174}] \textbf{Collection of ML Projects}
    
A total of 110 projects matched our selection criteria and were retained as subjects for our study. In order to ensure that no fork projects are present among our studied systems, and that at least one ML library mentioned in step \circled{1} are covered in our dataset, the first author of this paper manually examined the repository of each of the 110 projects, and all 110 study projects meet these criteria. 

% \Foutse{what is the outcome of this step? did you exclude any project? please provide precise information!!!} %In \autoref{tab:table1}, we present a descriptive statistic of the projects selected for our study.  %presents a statistical description of the selected projects under different metrics.

\autoref{tab:table1} presents an overview of our selected projects. %In total, we analyzed 110 open-source projects with an average of 53,579 sources lines of code. 
Line \textbf{SLOC} present a summary statistic of source lines of code, ranging from the mean to the max sources lines of code for our studied subjects.
Line \textbf{\#Commits} is the summary statistic for the number of commits.
Line \textbf{\#PythonFiles} denotes the summary statistic of the number of Python files present in our subjects at the time of analysis.
The summary statistics of the number of contributors in each project is presented at line \textbf{\#Authors}. Number of stars is an indicator of the popularity of a project, line \textbf{\#Stars} is a statistical description of the number of stars for our subjects with an average of 1,167 numbers of stars.
\begin{table}
  \centering
  \caption{Overview of studied subjects}
    \begin{tabularx}{1.1\textwidth}{lllllll}
    \toprule
    \textbf{Metric} & \textbf{Mean} & \textbf{Min} & \textbf{25th Quartile} & \textbf{Median} & \textbf{75th Quartile} & \textbf{Max} \\
    \midrule
    SLOC  & 53,579 & 244   & 3,647 & 10,662 & 31,228 & 2,010,902 \\
    \midrule
    \# Commits & 759 & 42 & 172 & 323 & 782 & 9,139 \\
    \midrule
    \# PythonFiles & 272 & 5 & 35 & 101 & 210 & 7,510 \\
    \midrule
    \# Authors & 18 & 2 & 3 & 5 & 13& 261 \\
    \midrule
    \# Stars & 1,167 & 1& 2 & 20& 284 & 32,770 \\
    \bottomrule
    \end{tabularx}%
  \label{tab:table1}%
\end{table}%
    \item[\ding{175}] \textbf{Identification of logging libraries contained in the projects %Qualitative Analysis
    }
    
To obtain information about each logging statement, we built a custom Abstract Syntax Tree (AST) parser from the Python built-in package AST\footnote{ \url{https://bit.ly/3gxsOkc}} and used it to extract all import statements contained in all the Python files of our selected projects. We extracted a total of 13,270 \texttt{import} statements (Steps \circled{3} - \circled{4}, \autoref{fig:researchprocess}). At step \circled{4}, we manually analyzed the extracted \texttt{import} statements to identify the logging libraries involved. We identified the following 12 libraries in the projects of our dataset: %In total, we identified 13 %among others the following 
%libraries; i.e., 
logging\footnote{ \url{https://bit.ly/3gxsOkc}}, rospy\footnote{ \url{https://bit.ly/3shluMk}}, hparams\footnote{ \url{https://bit.ly/3sdKpk2}}, logger\footnote{ \url{https://bit.ly/3VMYr9K}}, warnings\footnote{ \url{https://bit.ly/3CVpxCR}}, callbacks\footnote{ \url{https://bit.ly/3SozpLk}}, wandb\footnote{ \url{https://bit.ly/3Sod80f}}, tensorflow\footnote{ \url{https://bit.ly/3guvYoR}}, caffe.log\footnote{ \url{https://bit.ly/2QjqYSf}}, dllogger\footnote{ \url{https://bit.ly/3TJyW7m}}, ml-logger\footnote{ \url{https://bit.ly/3z0Xz7O}}, mlflow\footnote{ \url{https://bit.ly/2KiwDFd}}.
    \item[\ding{176}] \textbf{Extraction of Logging Statements}

To identify logging statements contained in each file, the following steps were taken: (i) For each Python file in the 110 projects, we use the AST module from Python 3 to generate an AST for the file. (ii) We use AST Python libraries to extract the function call related to the 13 logging libraries mention above. (iii) We analyzed each of the libraries mentioned in the previous step, then identified regular expressions that could match a logging statement of these libraries. (iv) Accordingly, we used the following expressions: log, info, debug, error, fatal, warn, callbacks, warnings, exception, and critical as a visitor pattern\footnote{\url{https://bit.ly/3TKBeD8}} to get all the function calls corresponding to a logging statement. In order to minimize false positives in our logging statement dataset, we manually removed expressions containing \enquote{log} that are not logging statements (e.g., \enquote{dialog}, \enquote{login}, \enquote{numpy.log}, \enquote{tensor.log}, etc) and we also removed expressions corresponding to logging configuration (e.g., \enquote{caffe.init\_log()}, \enquote{logging.getLogger}, \enquote{logging.disable}, \enquote{logging.config}, etc) and this allowed us to keep a clean dataset\footnote{\url{https://bit.ly/3DlmGow}}. This logging statement extraction process is inspired by the work of \citep{chen2017characterizing,zhai2019test}. We obtained a total of %with 
$\sim$21K logging statements. The results for step \circled{5} answer our research question \textbf{RQ1}.

    \item[\ding{177}] \textbf{Stratified Random Sampling}

In order to answer research questions \textbf{RQ2} and \textbf{RQ3}, we prepared our dataset for qualitative analysis using stratified random sampling. We opted for a stratified sampling approach to ensure a proportionate representation of projects in our sample. %; i.e.,  with a high proportion of logging statements are more representative in the sample. 
We sampled a total of $\sim$2K logging statements, which corresponds to more than 95\% of confidence level with a 5\% confidence interval \citep{yamane1967statistics}.

\subsection{Data analysis}
\label{analysis}
   % \item[\ding{178}] \textbf{Qualitative Analysis}
    % \Foutse{please separate data collection from data analysis!!! you are mixing the two! this step is part of the analysis method, please move it to a different section!}}
    
We perform a manual analysis of the logging statements sampled at step 6, categorizing the logging statements and mapping them to the different phases of the ML pipeline (step 7 on \autoref{fig:researchprocess}). Specifically, %  each statement to manuallyIn this step, we perform qualitative analysis aimed at identifying different type of logging statement and the ML pipeline where logging statement are prevalent. For 
for each logging statement found in our sample, we identified in the Python file and the specific class and function where it appears. Next, we identify the phase of the ML pipeline \citep{amershi2019software}, that is concerned by the implementation contained in the identified file (including the logging statement). To map logging statements to ML pipeline phases, we attribute a label to each phase of the ML pipeline and assign to each identified logging statement, the label corresponding to its phase (as identified previously). This allows us to answer  %  corresponding of the phase concerned by the logging steto each phase of the ML pipeline to the corresponding logging statement.   and in which class and function it is used, then identify in which pipeline it should be assigned 
\circled{\scalebox{0.5}{RQ2}}. The manual analysis and the labeling were performed in parallel by two authors, and the third author acted as a referee for ambiguous cases where opinions diverged. All the diverging cases were discussed until reaching a consensus. To understand why developers %A second analysis was performed to understand why AI practitioners 
use logging statements in ML-based applications and answer our research question \circled{\scalebox{0.5}{RQ3}}, we manually analyzed the source code where each logging statement is placed in order to identify the reason why ML practitioners inserted these logging statements in their project.
\end{enumerate}

\section{Case Study Results}
\label{sec:section3}
In this section, we present the results of our three research questions. For each research question, we describe the motivation, the approach followed to answer the research question and our obtained results. 

\subsection*{\textbf{RQ1:} What are the characteristics of ML-based application logging practices?}
\label{sec:section3.1}

\subsubsection*{Motivation}
\label{sec:section3.1.1}
Many research studies have been conducted to characterize logging practices in traditional software application \citep{chen2017characterizing, yuan2012characterizing, zeng2019studying}, showing the importance of logging in software systems. The conclusion from these researches shows that, logging practice in traditional applications is widely used. Many research works have used AI algorithms to make prediction and anomaly detection based on log files \citep{nagaraj2012structured, xu2009detecting}. In addition, research has been done to help developers in their logging decisions \citep{zhao2017log20, zhu2015learning}. However, none of these research efforts have focused on the characteristics of logging in ML applications.
Intuitively, there exists significant difference in the programming paradigms between ML and traditional application development: ML practitioners use a \textit{data driven} approach, which lets them develop and train models with a set of parameters and hyperparameters, on large data, by contrast traditional software developers use a \textit{logic driven} approach which directly implements the program logic. Thus, intuitive, the logging of ML-based applications would be different from that of traditional applications. Therefore, in this research question, we want to study the characteristics of logging practices in ML-based applications.

\subsubsection*{Approach} 
\label{sec:section3.1.2}
Similar to previous studies on the logging practices of traditional software applications~\citep{chen2017characterizing, yuan2012characterizing, zeng2019studying, zhu2015learning}, we have studied the following aspects of the characteristics of logging practices in ML-based applications.
\begin{itemize}
    \item[$-$] \textbf{Logging libraries types.} During the data gathering process, we obtained a dataset of \texttt{import} statements, from which we performed a qualitative analysis in order to identify logging libraries. We also make a distinction between ad-hoc logging for Al-based applications, and general-purpose logging libraries.  To achieve this task, the three authors have performed analysis of the \texttt{import} dataset in order to identify potential logging libraries. We search online materials (e.g., through the GOOGLE \footnote{\url{https://bit.ly/3VPCmaM}} search engine) in order to have more information about suspected logging libraries. Once the libraries were identified, we did additional research to find out more information about their usage. The goal of this investigation is to identify patterns that can be used further to extract the associated logging statements. \citep{chen2017characterizing} describe general-purpose logging libraries as libraries that developers can define a logging level, i.e., INFO, DEBUG, WARN, ERROR, FATAL, EXCEPTION, CRITICAL. Then Ad-hoc logging libraries as libraries designed for the purpose of logging particular information which will not be or tangled in case of concurrent logging with other type of libraries. We have manually labeled our dataset of logging statements obtained during the data gathering according to two categories, general-purpose logging and ML-specific logging. The summary of logging libraries found is presented in \autoref{table:logginglibrairies}. 
% Please add the following required packages to your document preamble:
\begin{table}[]
\caption{Summary of logging libraries in ML-based application}
\begin{adjustbox}{width=\columnwidth}
\begin{tabular}{|l|l|l|l|}
\hline
\textbf{Logging   libraries type}            & \textbf{Libraries} & \textbf{release by}          & \textbf{main purpose}                                                                                                                              \\ \hline
\multirow{9}{*}{ML-specific   logging}        & hparams             & Python software   foundation & Use to log   hyperparameter of the model                                                       \\ \cline{2-4} 
& wandb               & Weight \& Biases             & \begin{tabular}[c]{@{}l@{}}Use   for dataset versioning, track hyperparameter,\\      Experiment tracking\end{tabular}                            \\ \cline{2-4} 
& TensorFlow          & Google                       & \begin{tabular}[c]{@{}l@{}}Use to log and summarize ML information when developing with\\       TensorFlow\end{tabular}                           \\ \cline{2-4} 
& logger              & NVIDIA                       & Use to log ML   information at inference stage                                                                                                     \\ \cline{2-4} 
& caffe.log           & Berkeley AI Research         & \begin{tabular}[c]{@{}l@{}}Use to log ML information when developing \\      with caffe\end{tabular}                                              \\ \cline{2-4} 
& mlflow              & Databricks                   & \begin{tabular}[c]{@{}l@{}}Use   to track ML experiment in training and inference stage, logging   metric\\      and hyperparameter\end{tabular} \\ \cline{2-4} 
& ml-logger           & Python software   foundation & Tracking   ML experiment                                                                                                                          \\ \cline{2-4} 
& dllogger            & NVIDIA                       & Tracking   ML experiment                                                                                                                          \\ \cline{2-4} 
& callbacks           & Keras                        & Tracking   ML experiment, do early stopping,                                                                                                       \\ \hline
\multirow{3}{*}{General   logging libraries} & logging             & Python software   foundation & Log   all kind of information in python                                                                                                            \\ \cline{2-4} 
& rospy               & Open robotics                & \begin{tabular}[c]{@{}l@{}}Use   to log all kind of information when working \\      in robot operating system with rospy libraries\end{tabular}  \\ \cline{2-4} 
& Warnings           & Python software   foundation & Use   to log warnings message in all kind of project in Python                                                                                     \\ \hline
\end{tabular}
\end{adjustbox}
\label{table:logginglibrairies}
\end{table}

    \item[$-$] \textbf{The density of logging statements.} We measure log density on the last version of each project at the time we perform this study, through the following steps: (i) after cloning the last version of a project, we use the cloc\footnote{\url{https://bit.ly/3DgYhjQ}} tool, which allows us to evaluate the total number of lines of code present in a project (SLOC), while excluding comments and empty lines. (ii) Then, we evaluated the number of logging statements present in each project by using a custom-built AST\footnote{\url{https://bit.ly/3F5bI7S}} parser \ref{sec:datacollection}, which allowed us to extract, then make a quantitative analysis of logging statements present in a given project. (iii) Finally, we calculated the number of lines of code per logging statement using the \( \frac{SLOC}{NL} \) formula. Such a metric has already been used in previous works \citep{yuan2012characterizing}. 
    \item[$-$] \textbf{The verbosity levels of logging statements.} \citep{li2017log} shows through their studies the importance of assigning the correct log level to a logging statement and the difficulty for developers to determine the appropriate log level. Similar to \citep{zeng2019studying} we study the distribution of logging statements in ML-based applications. From the logging statement dataset extracted during our study design process (Section \ref{sec:datacollection}), we evaluate the distribution of the verbosity levels based on the following patterns, i.e., info, warn, debug, error, except, critical, fatal, which are the verbosity levels suggested in the python's official documentation\footnote{\url{https://bit.ly/3TJSpEY}}.
\end{itemize}

We have studied the three aspects mentioned above on the 110 open-source projects, in order to understand the characteristics of logging practices in ML-based applications.

\subsubsection*{Results}\label{sec:section3.1.3}
\begin{figure}[htp]
\includegraphics[width=0.9\textwidth]{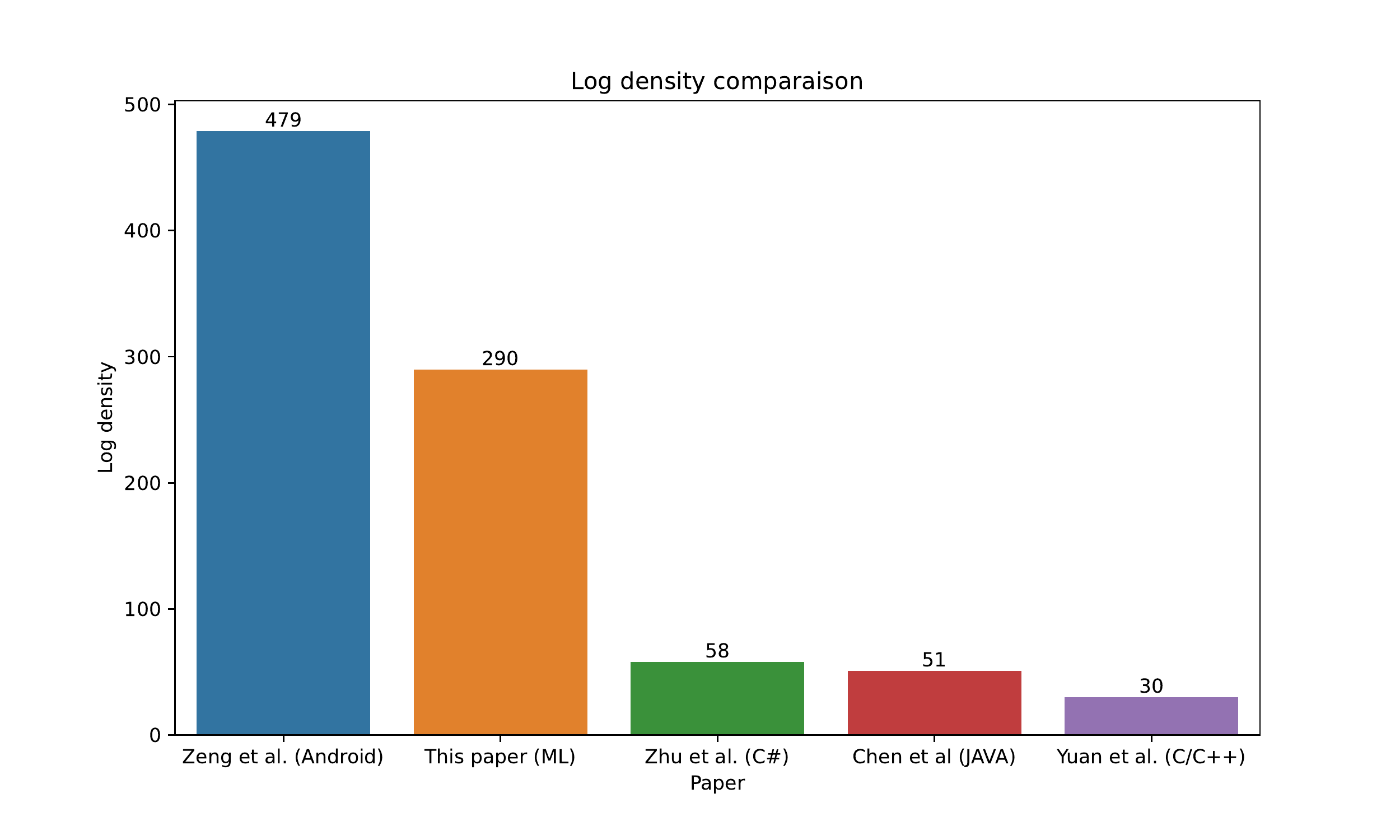}
\centering
\caption{Comparison of log density between ML-based applications studied in our paper and other types of applications studied in prior studies.} 
%  \Foutse{the values of density on this figure are strange!!!}
\label{fig:log_density_paper} 
\end{figure}

In this section, we present the results of our first research question, discuss our results, findings, and compare them to prior studies. We discover three main findings in our RQ1. 
\finding{\label{find: finding1} Logging statements in ML-based applications is commonly used but less pervasive than in JAVA, C\#, C/C++ applications and more pervasive than in Android applications.}
From the 110 studied projects, 78.18\% (86) contain at least one logging statement in the project. This shows that logging is prevalent in ML-based applications. Log density estimates the likelihood for a developer to insert logging statements in a project, the smaller
% \Foutse{smaller density? means fewer statements not the opposite!!! please check your statement. your figure 3 is unclear to me, what is the Y axis? a density is a ratio normally? i expected it to be number of loging statements per line of code, in that case it is strange to have values such as 400...please explain! clarify your computation! did you apply logarithm?} 
it is the more likely that we will find a logging statement in that project. We can observe from \autoref{fig:log_density_paper} that there is a very high probability that a developer writes a logging statement in a C/C++ project according to \citep{yuan2012characterizing} compare to others (on average one logging statement per thirty lines of code versus one logging statement per fifty-one line of code in JAVA application). For ML-based applications, the log density is smaller 
% \Foutse{it is larger 1/290 vs 1/479 if i follow your logic used to compare c/c++ and Java!!!}
compared to Android applications (on average one logging statement per two hundred and ninety lines of code vs one logging statement per four hundred and seventy-nine logging statement) but larger 
than C\# applications (290 vs 58). Thus, the probability for an ML practitioner introducing a logging statement in an ML application is higher than an Android developer introducing a logging statement in an Android application but lower than in JAVA, C\#, and C/C++ applications.
% \Foutse{i dont think so...please revise this section and clarify your computation of density!!!}

\begin{figure}[htp]
\includegraphics[width=0.8\textwidth]{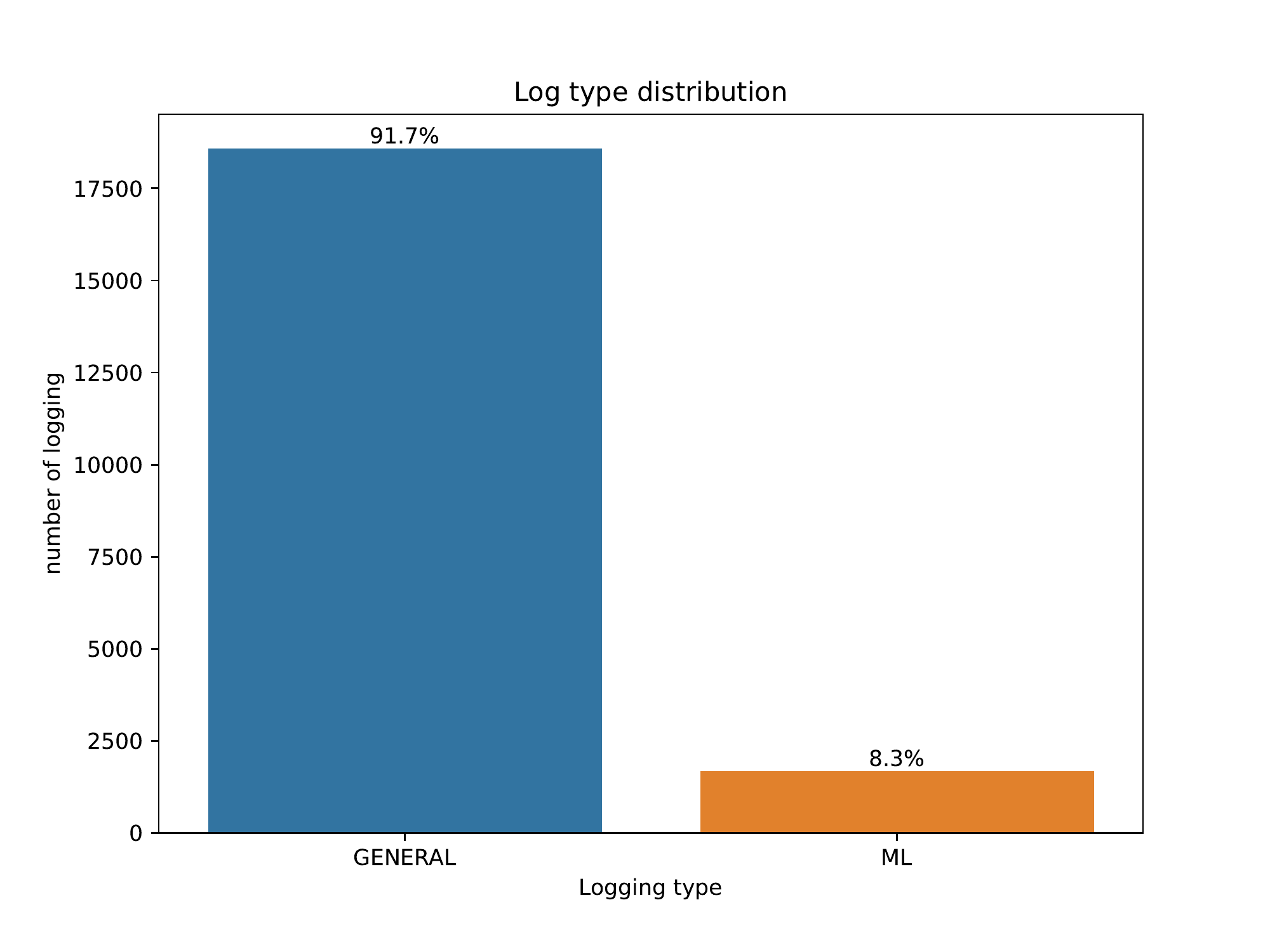}
\centering
\caption{Logging type distribution.} 
\label{fig:logging_distribution} 
\end{figure}

\finding{\label{find: finding2}The majority of logging statements in ML-based applications use general-purpose logging libraries.}
\autoref{fig:logging_distribution} shows the distribution of logging statements according to whether they belong to the general-purpose logging libraries or ad-hoc logging libraries for ML-based applications. The general-purpose logging statements are 11 times more used than the ML component-specific logging statements. This domination can be explained by the fact that: (i) general-purpose logging statements are used in both ML and non-ML components, while ML-specific logging libraries for ML-based applications are used only in the ML components. For example, general-purpose logging statements are used in the ML components (data processing), as presented in the following code snippet which can be found in AI-training\footnote{\url{https://bit.ly/3F1mdsy}}
% \Foutse{you are usually mixing AI and ML while these two are not exactly the same thing...please pick one term and use consistently!!! AI is more than ust ML} AI-training is the name of a project we can't change it.
\begin{lstlisting}[language=Python, caption= Code snippet of general logging statement used in ML component, captionpos=b, label={lst:loggingcode1}, basicstyle=\tiny]
def convert_ilsvrc2010(directory, output_directory,
                       output_filename='ilsvrc2010.hdf5',
                       shuffle_seed=config.default_seed):

    with h5py.File(output_path, 'w') as f:
        log.info('Creating HDF5 datasets...')
        prepare_hdf5_file(f, n_train, n_valid, n_test)
        log.info('Processing training set...')
        process_train_set(f, train, patch, n_train, wnid_map, shuffle_seed)
        log.info('Processing validation set...')
        process_other_set(f, 'valid', valid, patch, valid_groundtruth, n_train)
        log.info('Processing test set...')
        process_other_set(f, 'test', test, patch, test_groundtruth,
                          n_train + n_valid)
        log.info('Done.')
    return (output_path,)


def prepare_metadata(devkit_archive, test_groundtruth_path):

    # Ascertain the number of filenames to prepare appropriate sized
    # arrays.
    n_train = int(synsets['num_train_images'].sum())
    log.info('Training set: {} images'.format(n_train))
    log.info('Validation set: {} images'.format(len(valid_groundtruth)))
    log.info('Test set: {} images'.format(len(test_groundtruth)))
    n_total = n_train + len(valid_groundtruth) + len(test_groundtruth)
    log.info('Total (train/valid/test): {} images'.format(n_total))
    return n_train, valid_groundtruth, test_groundtruth, wnid_map
\end{lstlisting}
They are also used in the model training step, as in the following code snippet present in  cifar10-human-experiments\footnote{\url{https://bit.ly/3ShI2XW}},
\begin{lstlisting}[language=Python, caption= Code snippet of general logging statement used in model training, captionpos=b, label={lst:loggingcode2}, basicstyle=\tiny]
    if human_tune:
        logger.info('- epoch {}    c10h_train    : {:.4f} (acc: {:.4f}) | c10h_val    : {:.4f} (acc: {:.4f})'.format(
            epoch, train_loss_meter.avg, train_accuracy, loss_meter.avg, accuracy))
        logger.info('-            c10h_train_c10: {:.4f} (acc: {:.4f}) | c10h_val_c10: {:.4f} (acc: {:.4f})'.format(
            c10h_train_c10_loss_meter.avg, c10h_train_c10_accuracy, c10h_val_c10_loss_meter.avg, c10h_val_c10_accuracy))
    else:
        logger.info('Epoch {} Loss {:.4f} Accuracy {:.4f}'.format(
            epoch, loss_meter.avg, accuracy))

    elapsed = time.time() - start
    logger.info('Elapsed {:.2f}'.format(elapsed))

\end{lstlisting}
(ii) ML practitioners quite often use general-purpose logging statements in ML components instead of specific logging statements for ML components in their applications. 
This is the case for example in \autoref{lst:loggingcode1} where ML practitioners used a general-purpose logging statement to log the model parameters. The same logging task could have been done using the logging statement \enquote{\ji{wandb.log()}} which is specifically designed to log model parameters during model training. This is the case in the following code snippet, which can be found in BPNN\footnote{\url{https://bit.ly/3FbDQWC}}.
\begin{lstlisting}[language=Python, caption= Code snippet of ML-specific logging statement used in model training, captionpos=b, label={lst:loggingcode3}, basicstyle=\tiny]
 if USE_WANDB:
                wandb.log({"RMSE_val": np.sqrt(np.mean(val_losses)), "RMSE_training": np.sqrt(np.mean(losses))}, commit=(not args.val_ood))       

                    if USE_WANDB:
                        wandb.log({"RMSE_valOOD1": np.sqrt(np.mean(val_losses1))})       

                    print("root mean squared validation OOD1 error =", np.sqrt(np.mean(val_losses1)))

                    if USE_WANDB:
                        wandb.log({"RMSE_valOOD3": np.sqrt(np.mean(val_losses3))})       

                    if USE_WANDB:
                        wandb.log({"RMSE_valOOD4": np.sqrt(np.mean(val_losses4))}, commit=True)  
\end{lstlisting}

\begin{figure}[htp]
\includegraphics[width=0.8\textwidth]{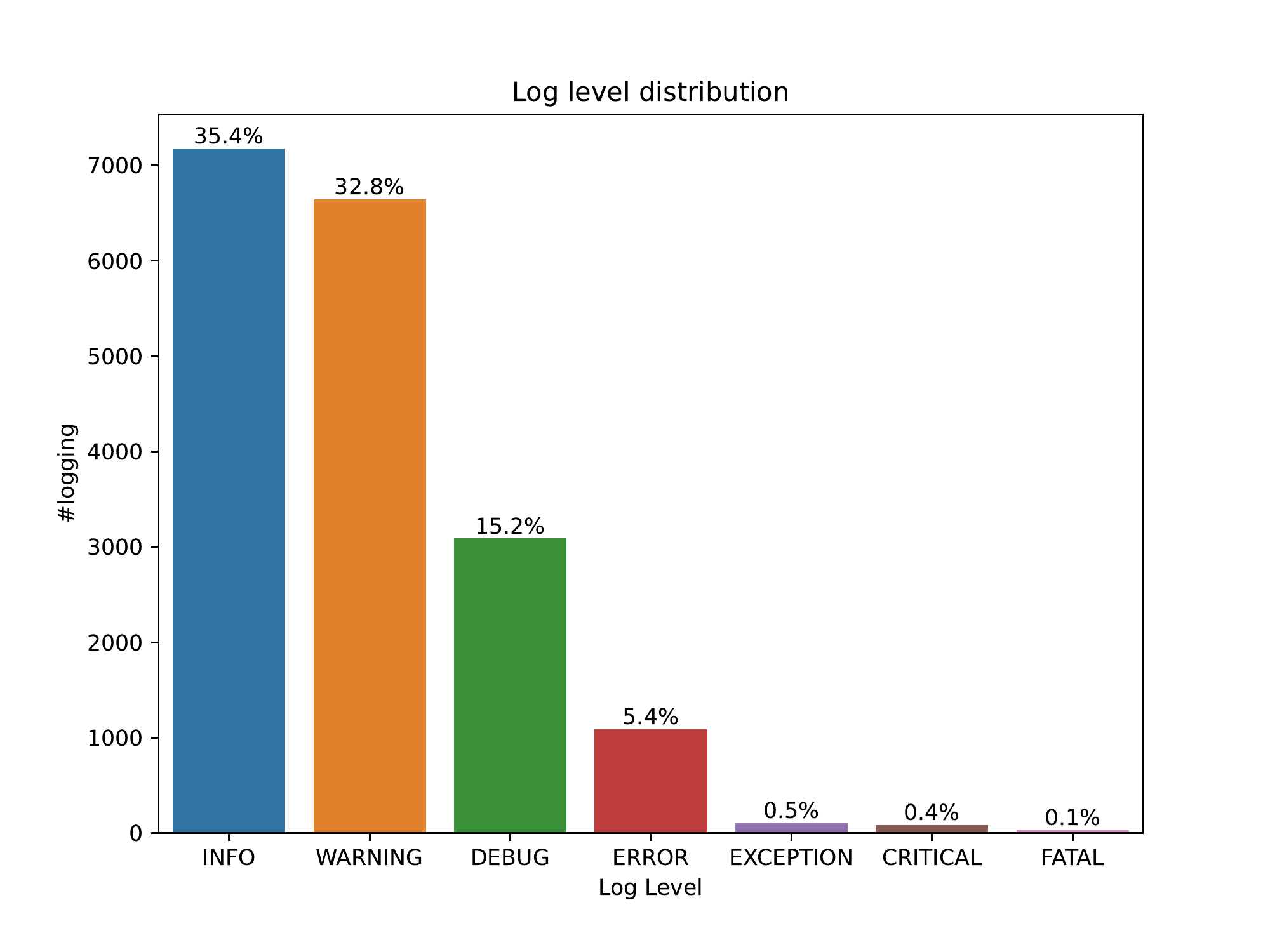}
\centering
\caption{Distributions of the logging statement levels.} 
\label{fig:logleveldist} 
\end{figure}

\finding{\label{find:logging_level_2}The majority of the logging statements in ML-based applications are in \ji{info} and \ji{warnings} levels, while \ji{info} level logging statements are the majority in C/C++ applications and \ji{debug}, \ji{error} level logging statements are the majority in Android Applications.}
\autoref{fig:logleveldist} shows the distribution of logging levels for the 110 studied projects. More than half of the logging statements in ML-based applications are in \ji{info} and \ji{warning} levels. This can be explained by the fact that ML practitioners often use these log levels in data and model management as shown in the examples \autoref{lst:loggingcode1}, \autoref{lst:loggingcode2}, where the info level is used for model and data management (recording model parameters and information during data processing). The warnings level is used to alert inconsistencies in data preprocessing operations. 
\begin{lstlisting}[language=Python, caption= Code snippet showing the usage of warning in data processing, captionpos=b, label={lst:loggingcode4}, basicstyle=\tiny]
def _contains_nan(a, nan_policy='propagate'):
        try:
            contains_nan = np.nan in set(a.ravel())
        except TypeError:
            # Don't know what to do. Fall back to omitting nan values and
            # issue a warning.
            contains_nan = False
            nan_policy = 'omit'
            warnings.warn("The input array could not be properly checked for nan "
                          "values. nan values will be ignored.", RuntimeWarning)
def kurtosistest(a, axis=0, nan_policy='propagate'):

    if n < 20:
        warnings.warn("kurtosistest only valid for n>=20 ... continuing "
                      "anyway, n=%i" % int(n))  

def dataset_from_dicts(self, dicts, indices=None, return_baskets=False, non_initial_token="X")
                    logger.warning(f"[Task: {task_name}] Could not convert labels to ids via label_list!"
                                   f"\nWe found a problem with labels {str(problematic_labels)}")
                # TODO change this when inference flag is implemented
                except KeyError:
                    logger.warning(f"[Task: {task_name}] Could not convert labels to ids via label_list!"
                                   "\nIf your are running in *inference* mode: Don't worry!"
                                   "\nIf you are running in *training* mode: Verify you are supplying a proper label list to your processor and check that labels in input data are correct.")
\end{lstlisting}
(Air-Pollution\footnote{\url{https://bit.ly/3SoY9Dh}}, farn\footnote{\url{https://bit.ly/3z0EMJy}}) \autoref{lst:loggingcode4} or model training \autoref{lst:loggingcode5} (DeepPavlov\footnote{\url{https://bit.ly/3TqKwVp}}) in ML component,
\begin{lstlisting}[language=Python, caption= Code snippet showing the usage of warning in model training, captionpos=b, label={lst:loggingcode5}, basicstyle=\tiny]
def train_evaluate_model_from_config() -> Dict[str, Dict[str, float]]:

    if 'train' not in config:
        log.warning('Train config is missing. Populating with default values')
    train_config = config.get('train')

    if 'evaluation_targets' not in train_config and ('validate_best' in train_config
                                                     or 'test_best' in train_config):
        log.warning('"validate_best" and "test_best" parameters are deprecated.'
                    ' Please, use "evaluation_targets" list instead')

    if iterator is not None:
        if to_validate is not None:
            if evaluation_targets is None:
                log.warning('"to_validate" parameter is deprecated and will be removed in future versions.'
                            ' Please, use "evaluation_targets" list instead')
                evaluation_targets = ['test']
                if to_validate:
                    evaluation_targets.append('valid')
            else:
                log.warning('Both "evaluation_targets" and "to_validate" parameters are specified.'
                            ' "to_validate" is deprecated and will be ignored')

    return res 
\end{lstlisting}
also its use in non-ML components as follows (Django\_web\footnote{\url{https://bit.ly/3F4EsgT}}).
% \newpage
\begin{lstlisting}[language=Python, caption= Code snippet showing the usage of warning in non-ML component, captionpos=b, label={lst:loggingcode6}, basicstyle=\tiny]
def methods_check(req, methods):
    try:
        assert req.method in methods
    except AssertionError:
        message = 'Method Not Allowed ({method}): {path}'.format(
            method=req.method, path=req.path)
        log.warning(message)
        raise ResponseNotAllowed(detail=message)
\end{lstlisting}

The distribution shown in Figure \ref{fig:logleveldist} is different from previous studies. The results of \citep{zeng2019studying} show that most logging statements in Android applications are at the \ji{debug} and \ji{error} level, while \citep{yuan2012characterizing} show that most logging statements in C/C++ applications are at the \ji{info} level.
% \vspace{50mm}
\begin{tcolorbox}[colback=black!4,colframe=black!50!white]
\textbf{Summary:} Logging practice in ML-based applications is different from logging practices in traditional applications. One of the major differences is the existence of multiple logging libraries in ML-based applications, which can be categorized into two groups: general-purpose and ML-specific logging libraries. General-purpose logging libraries are designed to be used mainly in traditional applications, and ML-specific logging libraries are designed to be used in ML-based applications. Our results show that general-purpose logging libraries are still the most used in ML-based applications, despite the existence of numerous ML-specific logging libraries. This finding suggests that ML practitioners might not be aware of the existence of ML-specific logging libraries, and might be adopting sub-optimal logging practices. The dominance of info and warning in logging levels can be explained by the fact that these log levels are most often used initially to log information throughout the application, i.e., in the ML and non-ML components of the application. These results are a preliminary step towards understanding logging practices in ML-based applications and call for further work to investigate good logging practices for ML-based applications.
\end{tcolorbox}

\subsection*{\textbf{RQ2:} Which phases of the ML pipeline are more prone to logging?}
\label{sec:section3.2}

\subsubsection*{Motivation}
\label{sec:section3.2.1}
%ML-based applications are designed using a data-driven approach, which consists of a pipeline that starts with reading data through training an ML algorithm on the data, then evaluating the model on test data until the model is deployed or put into production \citep{amershi2019software}. 
The results of our RQ1 indicate that ML practitioners widely use logging statements during the development of ML-based applications. This finding raises the question of log usage across the pipeline; i.e., which phases of the ML pipeline contain the most logging statements? Answering this question will help understand developers' needs for logging during the different phases of the ML pipeline, and help support the development of efficient logging libraries and recommenders for ML systems engineering. %,  ationsin which pipeline phases developers provide more attention to execution status and give directions for future research work.

\subsubsection*{Approach}
\label{sec:Section3.2.2}
%To answer the research question, we proceed as follows:
%\begin{enumerate}
    %\item Identifying ML pipeline phases.\\ 
    %A typical ML pipeline is composed of the following phases\citet{amershi2019software}: Model Training, Data Collection/loading, Data Processing, Model Evaluation/validation, and Model Deployment.  
    To select log statements for our analysis, we conduct a stratified random sampling as described in Subsection \ref{sec:datacollection}, and obtained $\sim$2K logging statements in our sample. %after having done a stratified random sampling on our logging dataset as mentioned in the . 
    Then, for each project in our sample, we analyzed the portion of the code where the logging statements are located and assigned each logging statement to one of the following five ML pipeline phases: Model Training, Data collection/loading, Data processing, Model evaluation, and Model deployment. 
   % \item Mapping logging statements to ML pipeline phases.\\
    We used the following clues to perform the matching of logging statements to the ML pipeline phases: (i) The path to the file containing a logging statement, often contains keywords such as \enquote{train}, \enquote{data}, \enquote{load}, \enquote{load\_data}, \enquote{model\_deploy}, \enquote{train\_image}, \enquote{fit}, \enquote{tokenizer}, \enquote{evaluator}, \enquote{data\_cleaning}. For example, it is the case for the paths of the following logging statements: pykale \footnote{ \url{https://bit.ly/3zioXy8}}, Air-Pollution \footnote{\url{https://bit.ly/3TXHbwE}},
    DeepLearningExamples \footnote{\url{https://bit.ly/3zlGQME}}.
    (ii) In the file, we pay particular attention to the name of the function where the logging statement is located, to the comment contained in the file, and those that surround the function. We also leverage the message written in the logging statement. For example, for this DeepLearningExamples\footnote{\url{https://bit.ly/3zoQXAd}} sample shown in the following Code Listing \ref{lst:loggingcode10}, we classify the logging statements in line \#2, \#3, \#7, \#16 as \enquote{model training} and those on lines \#21, \#22, \#24, \#30, \#31 as \enquote{model evaluation}. %%, which are summarizing in the following code snippet.
\begin{lstlisting}[language=Python, caption= Code snippet from training and evaluation phase, label={lst:loggingcode10}, captionpos=b,numbers=left, basicstyle=\tiny]
def train(n_token, cutoffs, rank, local_rank, num_core_per_host):
  tf.logging.info("num of batches {}".format(train_record_info["num_batch"]))
   tf.logging.info("step {} | lr {:8.9f} "
                        "| loss {:.2f} | pplx {:>7.2f}, bpc {:>7.4f}, tok/s {:>6.0f}".format(
                            curr_step, fetched[-2],
                            curr_loss, math.exp(curr_loss), curr_loss / math.log(2), throughput))
          dllogger_data = {
              'lr': fetched[-1],
              'train_loss': curr_loss,
              'train_perplexity': math.exp(curr_loss),
              'train_throughput': throughput,
          }
          dllogger.log(step=int(curr_step), data=dllogger_data)
          tf.logging.info("Model saved in path: {}".format(save_path))
    if rank == 0:
          tf.logging.info("Training throughput: {:>6.0f} tok/s".format(meters['train_throughput'].avg))
          
def evaluate(n_token, cutoffs):
  if FLAGS.max_eval_batch > 0:
      num_batch = FLAGS.max_eval_batch
  tf.logging.info("num of batches {}".format(num_batch))
  tf.logging.info("Evaluate {}".format(eval_ckpt_path))
  if (step+1) % (num_batch // 10) == 0:
        tf.logging.info(format_str.format(step+1, num_batch))
        dllogger_data = {
            'eval_latency': latency,
            'eval_throughput': throughput,
        }
        dllogger.log(step=step+1, data=dllogger_data)
    tf.logging.info("Evaluating with: bs {}, math {} ".format(FLAGS.eval_batch_size, "amp" if FLAGS.amp else "fp32"))
    tf.logging.info("| loss {:.2f} | pplx {:>7.2f}, bpc {:>7.4f}, tok/s {:>6.1f}, ms/batch {:>4.2f}".format(
        avg_loss, math.exp(avg_loss), avg_loss / math.log(2), meters['eval_throughput'].avg, meters['eval_latency'].avg))
\end{lstlisting}
%\end{enumerate}
   The manual classification task was performed by two authors with the kappa agreements of 0.71, i.e., substantial agreement. The ambiguous cases were discussed during a meeting. If an agreement couldn't be reached in the meeting, the logging statement was labeled as an %then we discussed ambiguous cases. If we couldn't reach an agreement, we added this to 
   unclassified case. In total, we were not able to classify 66 logging statements in our sample. We attribute this outcome to the complexity of the code containing the logging statements and--or the absence of the key elements listed above.
% Table generated by Excel2LaTeX from sheet 'ML-pipeline distribution'
\begin{table*}[hbt!]
  \centering
  \caption{Distribution of logging statement in the different phases of the ML pipeline}
  \begin{adjustbox}{width=\textwidth}
    \begin{tabular}{|p{8.285em}|p{17.145em}|p{30.285em}|r|}
    \toprule
    \textbf{ML pipeline} & \textbf{Description} & \textbf{Example} & \textbf{\% age} \\
    \midrule
    Model Training & Model training includes hyper-parameter tuning, optimizing cost function, fitting data, and model selection & logging.info('Training with a single process on 1 GPU.')\newline{}tf.logging.info("[*] num\_epochs: \%d" \% \_num\_epochs)\newline{}logging.info(f'Training time: {(elapsed / 60):.2f} minutes')\newline{}logger.info('Epoch[\%d] Time cost=\%.3f', epoch, (toc - tic))\newline{}logger.info(f"Loading pretrained files for: {', '.join(self.loadables)}")\newline{}dllogger.log(step=(epoch, steps\_per\_epoch, batch\_idx), data=dllogger\_data, verbosity=0)\newline{}dllogger.log(step="PARAMETER", data={'Scheduled\_epochs': num\_epochs}, verbosity=0) & 35.83\% \\
    \midrule
    Data Collection/loading & Data collection/loading  includes collecting and reading data  from different sources and  formats: online, CSV, Json, text, h5, SQL. & log.info('Recording driving data to \%s', self.hdf5\_dir)\newline{}logging.info('Load FTP fie without auth ({} fromm {})'.format(ftpURL, ftpFilePath))\newline{}logger.info("loading model from {}".format(self.path))\newline{}logger.info("Done loading the dataset") & 14.47\% \\
    \midrule
    Data Processing & Data processing step includes normalization, \newline{}transformation, validation and featurization of the data & logger.info(f"Moving label {repr(saved\_label)} from index "f"{index}, because {repr(label)} was put at its place.")\newline{}tf.logging.warning('`shuffle` is false, but the input data stream is ')\newline{}logger.info("Frequencies of rankings: {}".format(print\_dictionary(freq)))\newline{}logger.info(f"Resizing input dimensions of {type(self).\_\_name\_\_}"  f"from {old\_dims} to {new\_dims} to match language model") & 15.84\% \\
    \midrule
    Model Evaluation/validation & Evaluation of the model on validation or testing data, selection of the best model and validation of its correctness & logger.log("Save the best model into {:}".format(final\_best\_name))\newline{}logger.log("The best model has {:} weights.".format(best\_model.numel()))\newline{}logging.info(f'Latency Avg: {1000.0 * latency\_data.mean():.2f} ms')\newline{}tf.logging.info('Starting Evaluation.')\newline{}tf.logging.info("***** Running prediction*****")\newline{}logger.info('Final test accuracy = \%.1f\%\% (N=\%d)' \% (test\_accuracy * 100, len(test\_bottlenecks))) & 6.20\% \\
    \midrule
    Model Deployment & Deploying the model or the system in production & logger.info("deploying model " + self.args.triton\_model\_name +" in format " + self.lib.platform)\newline{}logger.info("model check failed with warning: [", error, "]")\newline{}logger.info("Warning during onnx.checker.check\_model in quantized model ignored")\newline{}LOGGER.debug("Existing deployed resources: \%s", existing\_resources) & 3.66\% \\
    \bottomrule
    \end{tabular}%
    \end{adjustbox}
  \label{tab:distribution}%
\end{table*}%
\subsubsection*{Results}\label{sec:section3.2.3}
In this section, we present the results of our second research question and highlight our findings.

\begin{figure}[htp]
\includegraphics[width=\textwidth]{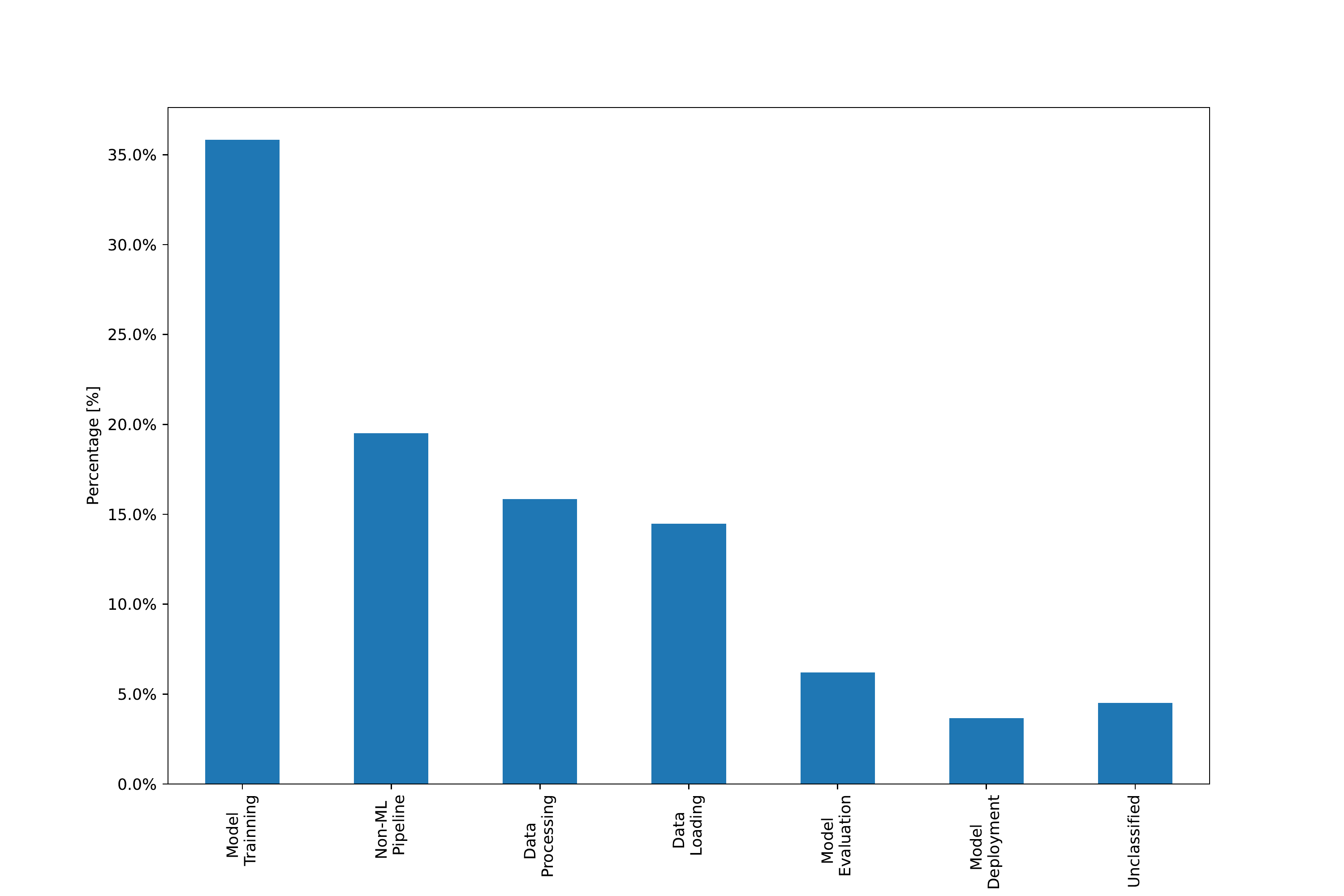}
\centering
\caption{Percentage of the analyzed sample of logging statements belonging to the different phases of the ML pipeline} 
\label{fig:logging_componentts} 
\end{figure}

From the $\sim$2K logging statements analyzed, 35.83\% of the logging statements belong to the model training phase and 19.51\% belong to the portion of code that isn't part of any ML component of the project.
% \heng{not shown in the table? yes because we want to focus in ML Pipeline but its show in the graph}. 
We also found that 15.84\%, 14.47\%, 6.20\%, and 3.66\% of logging statements belong respectively to Data Processing, Data Loading, Model Evaluation, and Model Deployment. We couldn't reach an agreement for 4.50\% (i.e., 66) of logging statements in our sample. \autoref{fig:logging_componentts} presents a distribution of the percentage of logging statements found in different components of the studied ML-based applications. One can see that the majority of logging statements are found in the model training component and that the model deployment phase contains the lowest percentage of logging statements. This result is not surprising since the model training phase produces a larger proportion of code than the other phases; there are many parameters to optimize in order to obtain a stable model and developers often have to log multiple pieces of information to guide their optimization process (training and re-training). In \autoref{tab:distribution}, we provide examples of logging statements that were found in our studied projects. The examples are grouped according to the ML pipeline phase to which they belong. In the description column, we present the characteristics of the phase, in particular, we list the elements that are considered during the classification of a logging statement into that phase. % presents a description of the ML pipeline and elements we considered in each pipeline.

\finding{\label{find:logging_effort_6}Overall, all five phases of the entire ML pipeline contain logging statements. However, the model training phase has the largest proportion of logging statements and the model deployment phase has the smallest proportion of logging statements.}

% \heng{RQ2 needs to be expanded: 1) show the distribution of log levels in each phase, which will provide an idea of what purpose of logging in each phase (e.g., debugging, warning, etc); 2) discuss the logging for each phase (some discussions on what are logged (using examples) in each phases and provide explanations). The two points can be combined. I suggest that use one paragraph for each phase.}

\subsection*{\textbf{RQ3:} Why do ML practitioners use logging?}
\label{sec:section3.4}

% \heng{My main concern is about the relationship between RQ2 and RQ3. RQ2 studies in which phases of the ML pipeline practitioners log. Part of RQ3 also studies the phases. Furthermore, the phases identified in RQ3 have differences from RQ2, which can cause confusion. I would suggest to add a paragraph in RQ3 approach to explain the difference between the phases in RQ2 and RQ3, why they are different, and why you need to study them in two separate RQs. In addition, in the results section, we need to map the phases in RQ3 to the phases in RQ2 to avoid confusion.}

\subsubsection*{Motivation}
\label{sec:section3.4.1}
In RQ1, we observed that logging statements are prevalent in ML-based applications but less than in traditional software applications. We also observed that ML developers conjointly use ML-specific logging libraries and general-purpose logging libraries. In this research question, we aim to get a better understanding of ML developers' need for logging, throughout the life cycle of ML-based applications. Specifically, we want to identify the type of information that is being logged at a specific phase of the ML pipeline. %the type of information that nee understand the root cause of these differences in logging practices between ML-based applications and traditional applications. 
A good understanding of logging practices in ML-based applications can help design efficient tools to support logging. It will also help devise guidelines for practitioners developing and maintaining ML-based applications. %-Because of this disparity between logging practices in ML-based applications and traditional applications, we wanted to understand the reasons behind the use of logging statements in ML-based applications. Understanding the reasons behind the use of logging statements in ML-based applications can guide researchers and practitioners to consider or design new logging tools suitable for ML-based software such as whylogs \footnote{\url{https://whylabs.ai/whylogs}} which is a very recent library for logging any type of data and tracking changes in the dataset throughout the ML-pipeline.

\subsubsection*{Approach}
\label{sec:section3.4.2}
We performed a qualitative analysis to understand the reasons for using logging statements in ML-based applications. We randomly sampled 380 logging statements from our dataset, which corresponds to a 95\% confidence level with 5\% confidence interval. We inspected each logging statement in detail, including the log level, the log text, and the variables in the logging statement, as well as the surrounding code. We combined this information with the clues mentioned in Section \ref{sec:Section3.2.2}, to determine the type(s) of information that has been logged and ML developers' reasons for logging this information.
% determine the reasons for the usage of the logging statement. Indeed, the analysis of the logging statement itself is indicative of what kind of information is being logged.
Next, we classified the logging statements based on ML developers' reasons for using them and the specific stage(s) of the ML pipeline at which the logging statements were introduced in the project. %We further scrutinized \Foutse{how exactly? be more specific} the logging information of each logging statement in order to summarize the types of information that has been logged. 
Each logging statement was scrutinized independently by the first and the third author of this paper. All ambiguous cases were discussed with the second author until a consensus was reached. %Since this qualitative analysis is subjective from one author to another, to mitigate this the first and third authors of this paper have performed this analysis independently, ambiguous cases were discussed with the third author until an agreement was reached.  

\subsubsection*{Results}
\label{sec:section3.4.3}
In this section, we report and discuss our findings about the types of information being logged in ML-based applications and the specific stages of the ML pipeline where the logging events occur. A stage is an important step in a phase of the ML pipeline. %, since we group those phase in five in RQ2. %our results obtained during the qualitative analysis to answer our third research question and highlight our findings.

\begin{figure}[htp]
\includegraphics[width=\textwidth]{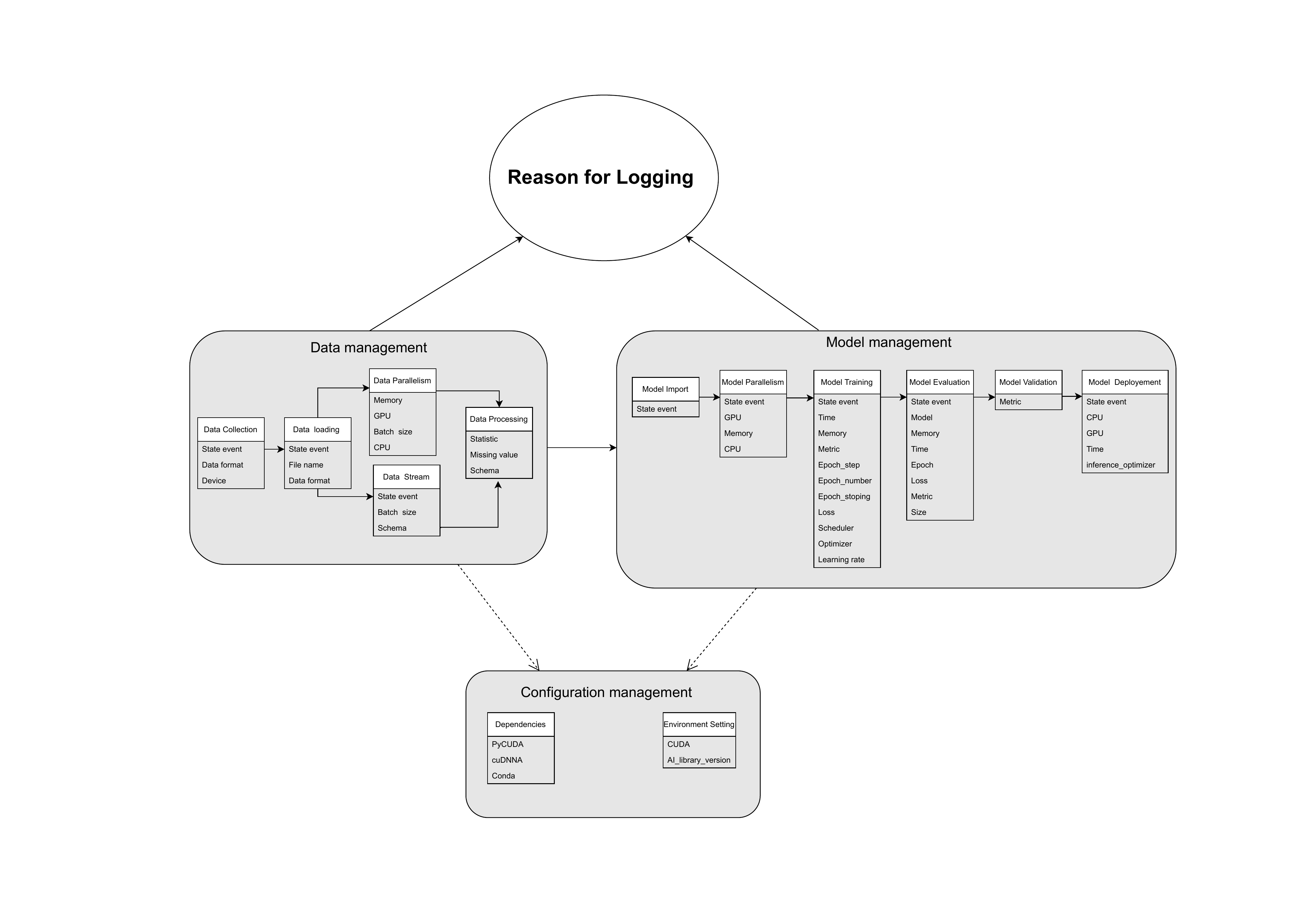}
\centering
\caption{Reason for Logging in ML-based application. \textit{Entity represent different stage in ML-based system. Attributes represent what type or what information is logged. }} 
\label{fig:purpose} 
\end{figure}

\textbf{Our analysis reveals two main reasons for logging in ML-based applications:  data management and model management}. We also identified a secondary reason which depends on the two previously mentioned reasons, namely configuration management.
% \heng{maybe name it configuration management?}.
\autoref{fig:purpose} summarizes our derived logging reasons. Each block of information in \autoref{fig:purpose} represents a reason for using logging in ML-based applications. Each entity in data management and model management represents a key stage (i.e., an important step in a phase of the ML pipeline), and the attributes are the types of information being logged. %recorded during a specific step.  
General purpose management depends on data management and model management entities because it allows to log information to ensure that the entire environment related to the training process or the data processing process is properly set up. Each entity in the general-purpose block represents a specific purpose for using a general-purpose logging library, and the attributes are the types of information being logged.

In the following, we elaborate in more detail on each of the reasons for logging identified in our analysis.\\ %statements that are observed by our study.
%\Foutse{I'm here!}\\
\noindent\textbf{Data management} concerns the lifecycle of data, from the collection (e.g., via APIs, web scraping, S3, and devices such as cameras, and phones) to the ingestion into the training process which includes the following steps: %. the  until its use to train an ML model through important steps which are, 
data loading, data parallelism (i.e., the distribution of the data across different nodes, which operate on the data in parallel), data processing, and data streaming. Throughout the lifecycle of the input dataset, ML practitioners often use logging to extract important information (for example about errors) and track the evolution of the data. %for traceability during the different stages of data manipulation and to track the evolution of data during its lifecycle. 
We have identified five stages in which logging is prevalent. % during which ML practitioners use log relevant information.
\begin{itemize}
    \item \textbf{Data Collection}: When collecting data, ML practitioners often log the following information of which some examples are presented in Listing~\ref{code:c-code_step1}: 
    % \Foutse{please describe the type of information being logged...} % sThis step is the very first one when ML practitioners want to build an ML-based application. It is a process of gathering qualitative or quantitative information through different techniques such as: interviews, observation, surveys, web scraping or social media monitoring. Some techniques require automation through scripts or programs in order to gather data and then structure it in a suitable format for ML models. During this process, a lot of information will be logged to make sure everything is done correctly. 
    % The attributes presented in the \enquote{Data Collection} entity in \autoref{fig:purpose} is a summary of the type of information usually logged by ML practitioners during data collection. 
    % Listing \autoref{code:c-code_step1} presents examples of logging statements identified at the data collection stage. Typically, ML  log 
    (i) \textbf{State event log} -which contains information about a particular event, e.g., recording data, processing erroneous data  during data collection. Those state event logs are introduced at the beginning, during, or at the end of those events, e.g., Line 1, 5, 6, 7 in Listing~\ref{code:c-code_step1}. 
    (ii) \textbf{Data format log} – the information logged is related to the type or format of data collected, e.g., Line 4, 1. (iii)
    \textbf{Device log} – the logged information is related to the data acquisition tools or devices, e.g., Line 2, 8. Logging statements present in Listing \autoref{code:c-code_step1} can be identified in the following projects: DeepLearningExamples\footnote{\url{https://bit.ly/3GDstYo}}, DeepCamera\footnote{\url{https://bit.ly/3XvpQ10}}.
    \begin{center}
    \begin{code}
    \begin{minted}[frame=lines,
    framesep=2mm,
    baselinestretch=1,
    bgcolor=LightGray,
    fontsize=\footnotesize,
    linenos]{python}
    log.info('Recording driving data to %s', self.hdf5_dir)
    log.debug('Grabbing images over ZMQ from %s', conn_string)
    log.debug('obz_exists? %r should_record? %r', obz is not None,self.should_record)
    logging.info("Converted {} frames in {}".format(num_frames, tfrecord))
    logging.info("Processed {} records".format(len(tfrecords)))
    logging.info("Corrupted record {}".format(tfrecord))
    logging.info("Processing record # {}: {}".format(record_id, record))
    logging.info('Camera: using Jetson onboard camera')
    
    \end{minted}
    \captionof{listing}{Logging statement examples in the data collection step}
    \label{code:c-code_step1}
    \end{code}
    \end{center}
    \item \textbf{Data loading}: Once the data collection is done, in this step the data is loaded from different sources such as: S3, directory, hard disk, in different formats such as JSON, CSV, etc. During this step ML practitioners usually record : (i) \textbf{State event log} - through this log here ML practitioners usually use state event log to monitor the data loading process through the records on the beginning of the loading process (e.g. line 4, 6), the errors occurred during the data loading (e.g., line 1), information about the file (e.g., line 4, 3) and the end of the data loading (e.g., line 5, 2). (ii) \textbf{Data information log} - here logged information is related to file name and the data formats as mentioned in \autoref{fig:purpose}. 
    Listing \autoref{code:c-code_step2} gives some examples of logging statements found in our study subjects, especially in the projects: cs-ranking\footnote{\url{https://bit.ly/3XCGFXQ}}, deepdrive\footnote{\url{https://bit.ly/3U1yhy1}} 
    \begin{center}
    \begin{code}
    \begin{minted}[frame=lines,
    framesep=2mm,
    baselinestretch=1,
    bgcolor=LightGray,
    fontsize=\footnotesize,
    linenos]{python}
    log.error('Could not load %s - skipping - error was: %r', h5_filename, e)
    log.info('finished loading %s', h5_filename)
    log.info('data name %s', dataset)
    logging.info('loading csv')
    logger.info("Done loading the dataset")
    log.info('loading %s', h5_filename)
    \end{minted}
    \captionof{listing}{Logging statement examples in the data loading step}
    \label{code:c-code_step2}
    \end{code}
    \end{center}
    \item \textbf{Data parallelism}: This step occurs when the collected data is too heavy compared to the available resources such as GPU and memory to be loaded directly. Then ML practitioners will perform data parallelization which consists in dividing the data according to the number of available GPUs or CPUs in order to accelerate the data processing and eventually the training process. They usually use: (i) \textbf{GPU/CPU log } - which consists of logging resource information related to the number of GPU or CPU used (e.g., line 3, 4). (ii) \textbf{Memory consumption log } -  are logging statement related to the memory utilization during the training process by ML practitioners (e.g., line 1, 2). (ii) \textbf{Batch log } - Here, data are loaded in batches, so the batch details like the batch size are logged (e.g., line 5). Some examples of logging statements that we have identified in our study subjects are presented in Listing \autoref{code:c-code_step3} and can be found in projects cs-ranking\footnote{\url{https://bit.ly/3Vkp0lz}}, rtg\footnote{\url{https://bit.ly/3EXYwRL}}.
    \begin{center}
    \begin{code}
    \begin{minted}[frame=lines,
    framesep=2mm,
    baselinestretch=1,
    bgcolor=LightGray,
    fontsize=\footnotesize,
    linenos]{python}
    log.warning(f"Going to buffer data; this may consume all the memory crash. Current usage={max_RSS()[1]}.")
    log.warning(f"Buffered {self.n_inp_recs:,} records Current memory usage={max_RSS()[1]}")
    logging.info("device_ids%s", gpu)
    logger.info("rank%s", cpu)
    logging.info("sampler size%s", batch_size)
    \end{minted}
    \captionof{listing}{Logging statement examples in the data parallelism step}
    \label{code:c-code_step3}
    \end{code}
    \end{center}
    \item \textbf{Data streaming}: This step occurs in online learning ML systems where the data is fed into ML algorithms in sequential order. This type of system is usually implemented when it becomes impossible to train the entire data set in a ML model, or when the data arrives in a streaming manner. During this stage, it is important for ML practitioners to ensure the consistency between different streams of data received by ML algorithms (e.g., line 2) we call this  (i) \textbf{Schema log }.  To this end it is important for ML practitioners to have information about the size of each stream of data received, their schema, monitor the process of receiving each stream of data through (ii) \textbf{State events log} (e.g., line 1, 2, 3, 4, 5). all this information will allow ML practitioners to have a track in case the distribution of the data changes. %During our analysis we have identified some logging statement 
    Listing \autoref{code:c-code_step4} present some of the relevant logging statements found in the projects: DeepCamera \footnote{\url{https://bit.ly/3XE6ren}}, attention-lvcsr\footnote{\url{https://bit.ly/3XlSuBD}}, and 
    attention-lvcsr\footnote{\url{https://bit.ly/3i8ulOw}}.
    \begin{center}
    \begin{code}
    \begin{minted}[frame=lines,
    framesep=2mm,
    baselinestretch=1,
    bgcolor=LightGray,
    fontsize=\footnotesize,
    linenos]{python}
    logger.info("Monitoring on auxiliary data finished")
    logger.info('No changes in schema detected.')
    logger.info("sending StopIteration")
    logger.info("sending {} arrays".format(len(data)))
    logger.info("Monitoring on auxiliary data started")
    \end{minted}
    \captionof{listing}{Logging statement examples in the data streaming step}
    \label{code:c-code_step4}
    \end{code}
    \end{center}
    \item \textbf{Data processing}: This step involves manipulating attributes and labels of data to produce meaningful information from the data. During the handling process, ML practitioners usually record (Listing \autoref{code:c-code_step5}) : (i) \textbf{Missing information log} - which are related to missing values, missing labels (e.g., line 1, 6, 7). (ii) \textbf{Statistic log} - this logging statement contains local estimators (mean, median, minimum, maximum), and variable estimators (variance, standard deviation) (e.g., line 3, 4). (iii) \textbf{Schema log} - logging statement here contains information about the schema of some data attribute (e.g., line 5). These examples can be found in the following subjects:
    csrank \footnote{\url{https://bit.ly/3Ox7tUZ}}, Air-Pollution.\footnote{\url{https://bit.ly/3Oz2svs}}
    \begin{center}
    \begin{code}
    \begin{minted}[frame=lines,
    framesep=2mm,
    baselinestretch=1,
    bgcolor=LightGray,
    fontsize=\footnotesize,
    linenos]{python}
    logger.info("Missing values {}: {}".format( col, np.isnan(arr).sum() / len(arr))
    logger.info("Sampled instances {} objects {}".format(X.shape[0], X.shape[1]))
    log.info(f"mean:{mean}, std:{std:.4f} ;; low:{low:.4f} high:{high:.4f} ;; invert:{invert}")
    logger.info("Min {}: Max {}".format(np.nanmin(arr), np.nanmax(arr)))
    logging.info('FORMAT --dateformat: {}'.format(FORMAT))
    tf.logging.fatal('Label does not exist %s.', label_name)
    logger.info("Missing values {}: {}".format( col, np.isnan(arr).sum() / len(arr))

    \end{minted}
    \captionof{listing}{Logging statement examples in the data processing step}
    \label{code:c-code_step5}
    \end{code}
    \end{center}
\end{itemize}

\noindent\textbf{Model management.}
Model management concerns the development of the model, which consists of developing, evaluating, validating and releasing it into production. A major difficulty of this stage is to track all the experiments performed in search of the best model and also better generalize on new data. ML practitioners often iterate on several experiments before arriving at the best model. Knowing which parameters or experiment led to the best model is a tedious task and can be time consuming, especially when done manually. One approach to solve this problem is to log all relevant information.  Logging all relevant information allows ML practitioners to reproduce or compare past experiments and then select the best model.
We have identified six important steps of model management during which ML practitioners  log relevant information.
\begin{itemize}
    \item \textbf{Model Import}: This step includes the import of the models or pre-trained weights, in different formats, during which the ML practitioners usually record \textbf{state events log} to ensure that the model importation process has been carried out correctly, such as the examples shown in Listing~\autoref{code:c-code_step6} from the AI-Training\footnote{\url{https://bit.ly/3Vg7BdM}} project.
    \begin{center}
    \begin{code}
    \begin{minted}[frame=lines,
    framesep=2mm,
    baselinestretch=1,
    bgcolor=LightGray,
    fontsize=\footnotesize,
    linenos]{python}
    logger.info(f"Loading pretrained files for: {', '.join(self.loadables)}")
    log.info('Downloading weights %s', folder)
    logging.info("Model loading done")
    logger.info("loading model from {}".format(self.path))
    \end{minted}
    \captionof{listing}{Logging statement examples in the model import step}
    \label{code:c-code_step6}
    \end{code}
    \end{center}
    \item \textbf{Model Parallelism}: This step occurs when there are memory constraints between the size of the model and the GPU or CPU device, as large models can't be trained on a single GPU. Thus, ML practitioners usually split the model onto several GPU devices.  During this step, ML practitioners usually record (i) \textbf{State event log} - which are logging statements containing information about the start and end of the training process (e.g., line 5). (ii) \textbf{Memory consumption log} - logging statements are related to memory consumed during the training stage. (iii)  \textbf{GPU/CPU log} - Information about the GPU and CPU devices used (e.g., lines 1, 2, 3, 6, 7). Listing \autoref{code:c-code_step7} shows some examples of such logging statements.
    \begin{center}
    \begin{code}
    \begin{minted}[frame=lines,
    framesep=2mm,
    baselinestretch=1,
    bgcolor=LightGray,
    fontsize=\footnotesize,
    linenos]{python}
    logging.info('Training in distributed mode with multiple processes, 1 GPU per
    process. Process %d, total %d.'% (args.rank, args.world_size))
    logger.info('device: {}'.format(device))
    logger.info("device: {} n_gpu: {} distributed training: {}".format(device, n_gpu, bool(args.local_rank != -1)))
    logger.info('begin training on multiple GPU')
    logger.info("device: {} n_gpu: {}".format(device, n_gpu))
    logging.info('Gradient averged for the rank of {}'.format(rank))
    \end{minted}
    \captionof{listing}{Logging statement examples in the model parallelism step}
    \label{code:c-code_step7}
    \end{code}
    \end{center}

    \item \textbf{Model Training}: In this step, ML practitioners use data to train ML models by manipulating many hyperparameters of the models and the optimization functions. This step is very time and resource-consuming and non-deterministic hence the need for ML practitioners to track a lot of information during this training process and record it through logging statements (e.g., examples shown in Listing \autoref{code:c-code_step8}). (i) \textbf{Time log} - Evaluation of training time or time taken to complete an epoch, e.g., Line 15. (ii) \textbf{Memory consumption log} - Evaluation of memory consumption, %will be performed and recorded 
    e.g., Line 14. (iii) \textbf{Hyperparameters log} - Recording of hyperparameters during the training process,  e.g., Line 2, 6, 7. (iv) \textbf{Metrics log} - Recording of metrics during training, e.g., Line 18, 13. (v) \textbf{Model size log} - Recording of the model size, e.g., : Line 4, 14, 16. (vi) \textbf{State events log} - Recording of states event which allows to track the beginning, the possible errors, the update of hyperparameters, the end of the training process e.g., Line 5, 8, 13, and eventually key information related to the end of the learning process of ML algorithms e.g., Line 1, 8. Some logging statements can be found in our following study subjects: BPNN \footnote{\url{https://bit.ly/3V2Y2PE}}, attention-lvcsr \footnote{\url{https://bit.ly/3EWJziG}}, AutoDL-Projects \footnote{\url{https://bit.ly/3tRdHpm}}, AutoDL-Projects \footnote{\url{https://bit.ly/3Xqnnoo}}.
    \begin{center}
    \begin{code}
    \begin{minted}[frame=lines,
    framesep=2mm,
    baselinestretch=1,
    bgcolor=LightGray,
    fontsize=\footnotesize,
    linenos]{python}
    logger.log("Early stop the pre-training at {:}".format(iepoch))
    log.info(f"New learning rate dividor = {self._learning_rate_cur_div}")
    log.info('pooling_0 shape: %s' % pooling_0.shape)
    logger.log("The base-model has {:} weights.".format(base_model.numel()))
    logger.log("[ONLINE] [{:03d}/{:03d}] loss={:.4f}, score={:.4f}".format(idx, len(env), future_loss.item(), score )
    logger.log("scheduler  : {:}".format(scheduler))
    logger.log("optimizer  : {:}".format(optimizer))
    logger.info("Initializing the training algorithm")
    log.info('Did not improve on the {} of {}'.format(m_name, self.score_best))
    logging.info('rank %s: failing epoch %s batch %s', rank, epoch, batch)
    wandb.log({"RMSE_training": np.sqrt(np.mean(losses))})
    logger.log("TRAIN [{:}] loss = {:.6f}".format(iter_str, loss.item()))
    logger.error("Error occured during training." + error_message)
    logger.log("FLOP = {:} MB, Param = {:} MB".format(flop, param))
    logging.info(f'Training time: {(elapsed / 60):.2f} minutes')
    logger.log("The model size is {:.4f} M".format(xmisc.count_parameters(model)))
    wandb.log({"RMSE_val": val_RMSE, "RMSE_training": training_RMSE}) 
    \end{minted}
    \captionof{listing}{Logging statement examples in the model training step}
    \label{code:c-code_step8}
    \end{code}
    \end{center}
    \item \textbf{Model Evaluation}: This step is part of the model development process. It includes the search for the best model that generalizes the data, as well as the evaluation of its performance on the validation or test data.  During this stage, ML practitioners usually log information that will allow tracing the whole evaluation process through: (i) \textbf{State events log} -  that will generally allow logging the beginning, the errors that occurred, and the end of the evaluation process,  e.g., Line 5. (ii) \textbf{Best model log} - log the best model information obtained during the fine-tuning phase,  e.g., Line 1. (iii) \textbf{Memory consumption log} - records the memory consumption during the evaluation process,  e.g., Line 2. (v) \textbf{Metric log} - records the performance measures (metric, loss) at each epoch,  e.g., Line 8, 3. (iv) \textbf{Model size log} - records the model size,  e.g., Line 7. \textbf{Inference time log} - records the time taken by the model to infer predictions,  e.g., Line 6. Listing \autoref{code:c-code_step9} are examples of logging statements collected in our study subjects at this stage, they can be found in AutoDL-Projects\footnote{\url{https://bit.ly/3GKwJ8J}}, DeepLearningExamples\footnote{\url{https://bit.ly/3VmRc7k}}.
    \begin{center}
    \begin{code}
    \begin{minted}[frame=lines,
    framesep=2mm,
    baselinestretch=1,
    bgcolor=LightGray,
    fontsize=\footnotesize,
    linenos]{python}
    logger.log("Save the best model into {:}".format(final_best_name))
    logger.log( ""Finish training/validation in {:} with Max-GPU-Memory of {:.2f} MB, and save final checkpoint into {:}"".format( convert_secs2time(epoch_time.sum, True)
    tf.logging.info('%s: Step %d: Validation accuracy = %.1f%% (N=%d)' %(datetime.now(),
     i, validation_accuracy * 100,len(validation_bottlenecks)))
    logger.info("***** Running evaluation *****")
    logger.info("time for inference {} perf {}".format(eval_end - eval_start, num_examples * 100 / (eval_end - eval_start)))
    logger.log("FLOP = {:} MB, Param = {:} MB".format(flop, param))
    logger.log("{:} {:} epoch={:03d}/{:03d} :: Train [loss={:.5f}, acc@1={:.2f}%, acc@5={:.2f}%] Valid [loss={:.5f}, acc@1={:.2f}%, acc@5={:.2f}%]".format( time_string(), need_time, epoch, total_epoch, train_loss, train_acc1, train_acc5, valid_loss, valid_acc1, valid_acc5,)
    \end{minted}
    \captionof{listing}{Logging statement examples in the model evaluation step}
    \label{code:c-code_step9}
    \end{code}
    \end{center}
    \item \textbf{Model Validation}:  The purpose of this step is to assess the precision and performance of the model obtained during the evaluation phase on real data.  To do so, ML practitioners record \textbf{metrics} used as a measure of performance for their ML system, as presented in Listing \autoref{code:c-code_step10}.
    \begin{center}
    \begin{code}
    \begin{minted}[frame=lines,
    framesep=2mm,
    baselinestretch=1,
    bgcolor=LightGray,
    fontsize=\footnotesize,
    linenos]{python}
    logger.info('Final test accuracy = %.1f%% (N=%d)' % (test_accuracy * 100, len(test_bottlenecks)))
    \end{minted}
    \captionof{listing}{Logging statement examples in the model validation step}
    \label{code:c-code_step10}
    \end{code}
    \end{center}
    \item \textbf{Model Deployment}: At this stage, in order to keep track of everything that happens during this phase ML practitioners often record information such as: (i) \textbf{State events log} - which are logging statements related to the process of converting the model into a lighter intermediate representation by applying graph optimizations, layer merging,  e.g., Line 1, 2, 3, 6. (ii) \textbf{GPU/CPU log} - Device information such as CPU and GPU are recorded during model prediction. (iii) \textbf{Time latency log} -  it's critical in a production environment to deliver inferences quickly. Hence, ML practitioners record the execution time of an inference,  e.g., Line 7. (iv) \textbf{Optimizer log} – there are different optimizers that can be used to convert an ML model into a lighter representation optimized for real-time inferences,  e.g., Line 4, 6. (iv) \textbf{Precision log} – the precision enabled by an optimizer is an important piece of information that ML practitioners record when the model is in production,  e.g., Line 5. Listing \autoref{code:c-code_step11} are examples of logging statements collected in our study subjects corresponding to the model deployment step.
    \begin{center}
    \begin{code}
    \begin{minted}[frame=lines,
    framesep=2mm,
    baselinestretch=1,
    bgcolor=LightGray,
    fontsize=\footnotesize,
    linenos]{python}
    logger.info(”model check failed with warning: [”, error, ”]”)
    logger.warning(”Warning during onnx.checker.check model in quantized model ignored”)
    logging.info('Total node count before and after TF-TRT conversion:', num_nodes, '->', len(frozen_graph.node))
    logger.info('TRT node count:',len([1 for n in frozen_graph.node if str(n.op)'TRTEngineOp']))
    tf.compat.v1.logging.info("Precision = %s", "fp16" if FLAGS.amp else "fp32")
    tf.compat.v1.logging.info('Converting graph using TensorFlow-TensorRT...')
    tf.compat.v1.logging.info("Total Inference Time W/O Overhead = %0.2f for Sentences = %d", predict_time_wo_overhead, num_sentences)
    \end{minted}
    \captionof{listing}{Logging statement examples in the model deployment step}
    \label{code:c-code_step11}
    \end{code}
    \end{center}
\end{itemize}

\noindent\textbf{Configuration management} steps are not explicitly represented in the ML pipeline but are essential to ensure that the ML components perform in a manner consistent with expectations over time. % concerns steps that we call invisible in the ML pipeline that depend heavily on data and model management. 
Configuration management activities generally include the management of configurations or dependencies on special devices and the management of libraries that are essential to ensure that ML components work as expected. Through our analysis, we have identified logs related to the following configuration management activities:
\begin{itemize}
    \item \textbf{Dependencies configuration log}: This activity is usually implemented when the ML component of the application will depend on a specific library in order to ensure that the ML module of an application can work properly. Hence, the need for ML practitioners to log information related to important libraries used. Listing \autoref{code:c-code_step12} Presents some logging statements found during this step. They can be found in our studied subjects: DeepLearningExamples\footnote{\url{https://bit.ly/3Eyc1WB}}, deepdrive\footnote{\url{https://bit.ly/3Ex9tZ5}}.
    \begin{center}
    \begin{code}
    \begin{minted}[frame=lines,
    framesep=2mm,
    baselinestretch=1,
    bgcolor=LightGray,
    fontsize=\footnotesize,
    linenos]{python}
    log.info('Installed UEPy python dependencies')
    warnings.warn('CVM does not support memory profile, using Stack VM.')
    warnings.warn("PyCUDA import failed in theano.misc.pycuda_init")
    logging.info("Using torch DistributedDataParallel. Install NVIDIA Apex for Apex DDP.")
    \end{minted}
    \captionof{listing}{Logging statement examples in the dependencies configuration step}
    \label{code:c-code_step12}
    \end{code}
    \end{center}
    \item \textbf{Environment setting log}: The environment in which ML applications run is different from traditional applications, in fact, ML applications generally need more resources or special devices such as CPU, and GPU in order to run properly. ML practitioners will generally ensure that the environment in which the application needs to run contains a minimum of resources. Therefore, information about critical resources will be logged (e.g., examples in Listing \autoref{code:c-code_step13}). Some examples can be found in the following subjects: DeepPavlov\footnote{\url{https://bit.ly/3TZ4mXi}}, AutoDL-Projects\footnote{\url{https://bit.ly/3tRawy3}}, AI-Training\footnote{\url{https://bit.ly/3ACmajS}}.
    \begin{center}
    \begin{code}
    \begin{minted}[frame=lines,
    framesep=2mm,
    baselinestretch=1,
    bgcolor=LightGray,
    fontsize=\footnotesize,
    linenos]{python}
    logger.log("cuDNN   Version  : {:}".format(torch.backends.cudnn.version()))
    logger.log("PyTorch Version  : {:}".format(torch.__version__))
    logger.warning("This recipe needs the sox-io backend of torchaudio")
    logger.log("CUDA available   : {:}".format(torch.cuda.is_available()))
    logger.log("CUDA GPU numbers : {:}".format(torch.cuda.device_count()))
    "logger.log(""CUDA_VISIBLE_DEVICES : {:}""
    .format( os.environ[""CUDA_VISIBLE_DEVICES""] if ""CUDA_VISIBLE_DEVICES"" in os.environ else ""None"")"
    caffe.log('Using devices %s' % str(gpus))
    _logger.warning("We are not able to detect the number of CPU cores." " We disable openmp by default.")
    _logger.info('Conda mkl is not available: %s', e)
    \end{minted}
    \captionof{listing}{Logging statement examples in the environment setting step}
    \label{code:c-code_step13}
    \end{code}
    \end{center}
\end{itemize}
\finding{
 ML practitioners use logging statements to record important information in \textit{data management} (e.g., recording the data format in data collection), \textit{model management} (recording the hyperparameters in model training), and \textit{configuration management} (e.g., recording the used resources or libraries). Our observations provide guidance for ML practitioners to improve their ML logging and insights for future efforts to improve ML logging practices (e.g., by providing ML logging libraries that facilitate the logging of different types of important information throughout the life-cycle of ML-based applications).
}

\section{Threats to validity}
\label{sec:threatsToValidity}
In this section, we discuss the potential threats to the validity of our research
methodology and findings.

\subsection*{\textit{External validity}}
The subjects used in this paper to answer our different research questions are open-source ML-based projects from GitHub. The selection of these projects may be subject to the following threats:

Our results and findings may not apply to ML projects written in other languages (e.g., JAVA, C\# or R) rather than Python since our analysis was mainly done on the portions of code written in Python in our study subjects. Hence, it is necessary that future works explore ML projects written in other programming languages since ML practitioners working with these languages may have different logging practices for ML projects. However, since Python is considered to be the \textit{lingua franca} for ML-based application development \citep{dilhara2021understanding}, we believe that our study provides a good understanding of the logging practices in ML-based systems. %, and we believe that our results can be used by numerous ML practitioners. 

\subsection*{\textit{Construct validity}} 
The threats to the construct validity of our research are related to errors that may have occurred during the extraction of logging statements. %  may result from the way we gather the data:

%The validity of our results firstly depends on how accurately we are able to detect and extract logging statements in our studied subjects. In this study, since our focus is on files written in Python, which is a dynamic programming language, a threat to the validity of our results is the extraction of 
To avoid extracting logging statements that have been commented out by developers, we have developed a  static code analyzer on top of the standard Python AST parser (which is widely used in the static analysis of Python code \citep{d2016collective, dilhara2021understanding}). This static code analyzer which is available in our replication package\footnote{Scripts and data files used in our research are available online and can be found here: \url{http://bitly.ws/yr6c}}library extract only uncommented statements; enabling us to avoid collecting logging statements that were commented out by developers. %, which allows us to avoid introducing commented statements when collecting logging statements. 

\subsection*{\textit{Internal validity}} 
We have manually mapped logging statements to their corresponding ML pipeline phase, to answer RQ2. Then, in RQ3, we manually analyzed the logging statements to identify the type of information logged and create a taxonomy. However, our manual analyses are subject to the subjective judgment of the people performing the analysis. This raises a threat to the internal validity of our results. To mitigate this threat, manual analyses were performed by the three authors of this paper with strong industry and academic backgrounds in ML systems engineering. Two authors performed the manual analysis, in case of disagreements we had a group discussion with the third author until a common agreement was reached. We believe that this approach reduces the chance of introducing false positives in our analyses. However, future replications and extensions of our work are desirable. All the data and scripts used in our study are available in our replication package\footnote{\url{http://bitly.ws/yr6c}}

\section{Related work}
\label{sec:relatedwork}

In this section, we introduce and discuss two areas of related works on logging practices: (i) research done on logging characteristics, and (ii) research done on logging decisions.  

\begin{enumerate}[label=(\roman*)]
    \item \textbf{Characterizing logging practice:} Many research works have been done in characterizing logging practices. \citep{yuan2012characterizing} conducts the first empirical study in the characterization of logging practices on four open-source projects written in C/C++ and found that logging is pervasive. \citep{chen2017characterizing} conducts a similar study by analyzing 21 projects written in JAVA which is a replication study whose objective is to generalize the findings obtained by \citep{yuan2012characterizing} for projects written in JAVA. Their result shows a difference in logging practices between applications written in JAVA and those written in C/C++. \citep{zeng2019studying} studied the logging practices in 1,444 Android projects and then compared their findings with those of previous works. However, none of the previous works have focused on logging practices in AI-based applications. Our paper fills this knowledge gap.
    \item \textbf{Logging decisions:} Logging decision involves research that helps developers decide where to introduce a logging statement and what level of verbosity should be assigned. These studies use AI algorithms to make predictions on where or what to log. \citep{ learntolog} propose a ``learning to log'' framework using machine learning techniques, which aims to help developers make decisions on where to add logging statements during development. Furthermore, \citet{zhenhaoase2020} used a deep learning-based approach to help developers in their logging decision at the block level.  \citep{li2017log} propose an ordinal regression model, which accurately provides the suggestion of the logging levels when developers add logging statement and \citep{9402068} suggest also log level by using Ordinal Based Neural Networks. More recently, \citep{mastropaolo2022using} propose LANCE a framework to generate a complete logging statement using deep learning. This obviously shows that AI algorithms are used to assist developers of traditional applications in their logging decisions but no research has been conducted to assist AI practitioners in their logging decisions in AI-based applications. Our work is therefore a good starting point for future research. 
\end{enumerate}

\section{Conclusions}
\label{sec:conclusion}
Logging practices have been adopted by developers as part of good programming practices. Logs generally allow developers to diagnose their programs at runtime in order to reduce maintenance efforts. Logging practices have been the subject of numerous studies in traditional software systems such as mobile applications, JAVA applications, and open-source applications. To the best of our knowledge, this paper presents the first attempt to study the practice of logging in ML-based applications. 
We studied 110 open-source ML-based applications from Github. Our research has yielded the following findings: Logging in ML-based applications is commonly used but less pervasive than in JAVA, C\#, C/C++ applications and more pervasive than in Android applications. The majority of logging statements are in \ji{INFO} and \ji{WARNING} levels. Moreover, we found that ML-based applications use two kinds of logging libraries: general logging libraries and those specific to ML-based applications. However, despite the existence of ML-specific libraries, general logging libraries remain the most used in ML applications. 
In order to identify which ML pipeline contains the most logging statements in ML-based applications, we performed a qualitative and quantitative analysis. Our findings indicate that the majority of logging statements are found in the \textit{model training} phase and the \textit{model deployment} phase contains the smallest portion of logging statements. Furthermore, %we have identified two main reasons why ML practitioners use logging statements in their applications. 
we observe that ML practitioners use logging statements to record important information related to data management (e.g., recording the data format in data collection), model management (recording the hyperparameters in model training), and configuration management (e.g., recording the used resources or libraries). 
The contribution of this paper is as follows:
\begin{itemize}
\item To the best of our knowledge, this is the first study that quantitatively and qualitatively analyzes the logging practice in ML-based applications.
\item We identified the general-purpose and ML-specific logging libraries that are used in ML-based applications.
%\item For the first time, a study revealed why logging statements are used in an ML-based application.
\item We identified ML phases in which practitioners use logs and the types of information they log. Our findings provide guidance for ML practitioners to improve their ML logging, as well as provide insights for future efforts to improve ML logging practices (e.g., by providing ML logging libraries that facilitate the logging of different types of information in different ML phases).
\end{itemize}

Overall, our findings highlight the need for more ML-specific libraries to support the development of ML-based applications. More research is also needed to improve our understanding of logging needs and challenges in the context of ML systems engineering. %on the logging practices of MLadvocate not only for the use of specific ML application logging libraries but also for the need for more ML application specific logging libraries. 

% \begin{acknowledgements}
% If you'd like to thank anyone, place your comments here
% and remove the percent signs.
% \end{acknowledgements}

% Authors must disclose all relationships or interests that 
% could have direct or potential influence or impart bias on 
% the work: 
%
\section*{Conflict of interest}

The authors declare that they have no conflict of interest.
\section*{Data availability statement}

The datasets generated during and/or analysed during the current study are available in the [foalem] repository, [\url{https://github.com/foalem/ML-logging-paper}].

% BibTeX users please use one of
%\bibliographystyle{spbasic}      % basic style, author-year citations
%\bibliographystyle{spmpsci}      % mathematics and physical sciences
%\bibliographystyle{spphys}       % APS-like style for physics
%\bibliography{}   % name your BibTeX data base

%%%%%%%%%%%%%%%%%%%%%%bibliography%%%%%%%%%%%%%%%%%%%%%%%%%%%%%%%%%%%%%%%%%%%%%%%

%%%%%%%%%%%%%%%%%%%%%%%%%%%%%%%%%%%%%%%%%%%%%%%%%%%%%%%%%%%%%%%%%%%%%%%%%%%%%%%%%
\bibliographystyle{spbasic}
\bibliography{reference}
\clearpage

\appendix
\begin{appendices}
\section{Figure}
\label{appendix}
\begin{figure}[h!]
\begin{subfigure}[b]{0.8\textwidth}
\includegraphics[width=\textwidth]{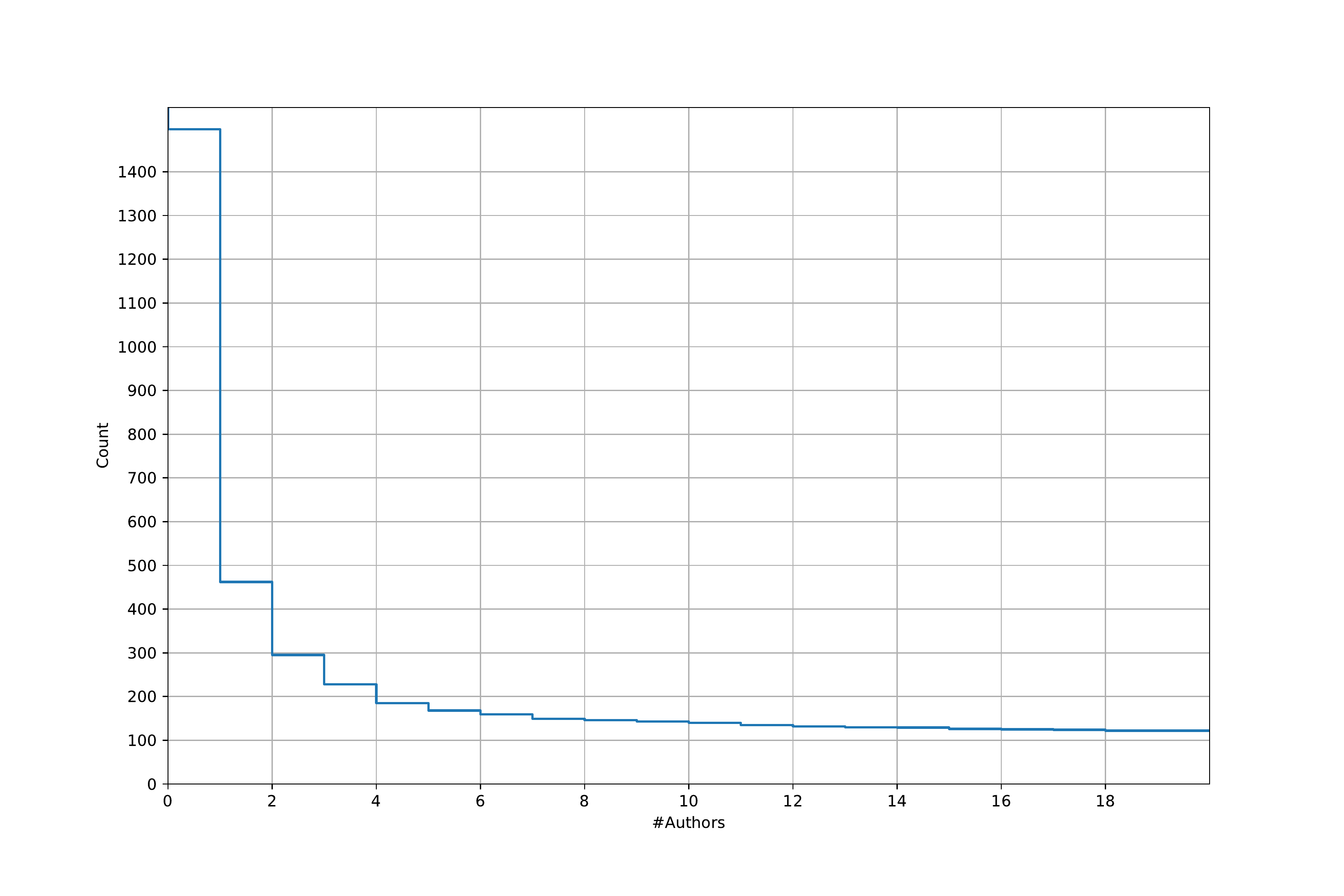}
\caption{Cumulative frequency curve base on \#Authors}
\label{img1}
\end{subfigure}
\hfill
\begin{subfigure}[b]{0.8\textwidth}
\includegraphics[width=\textwidth]{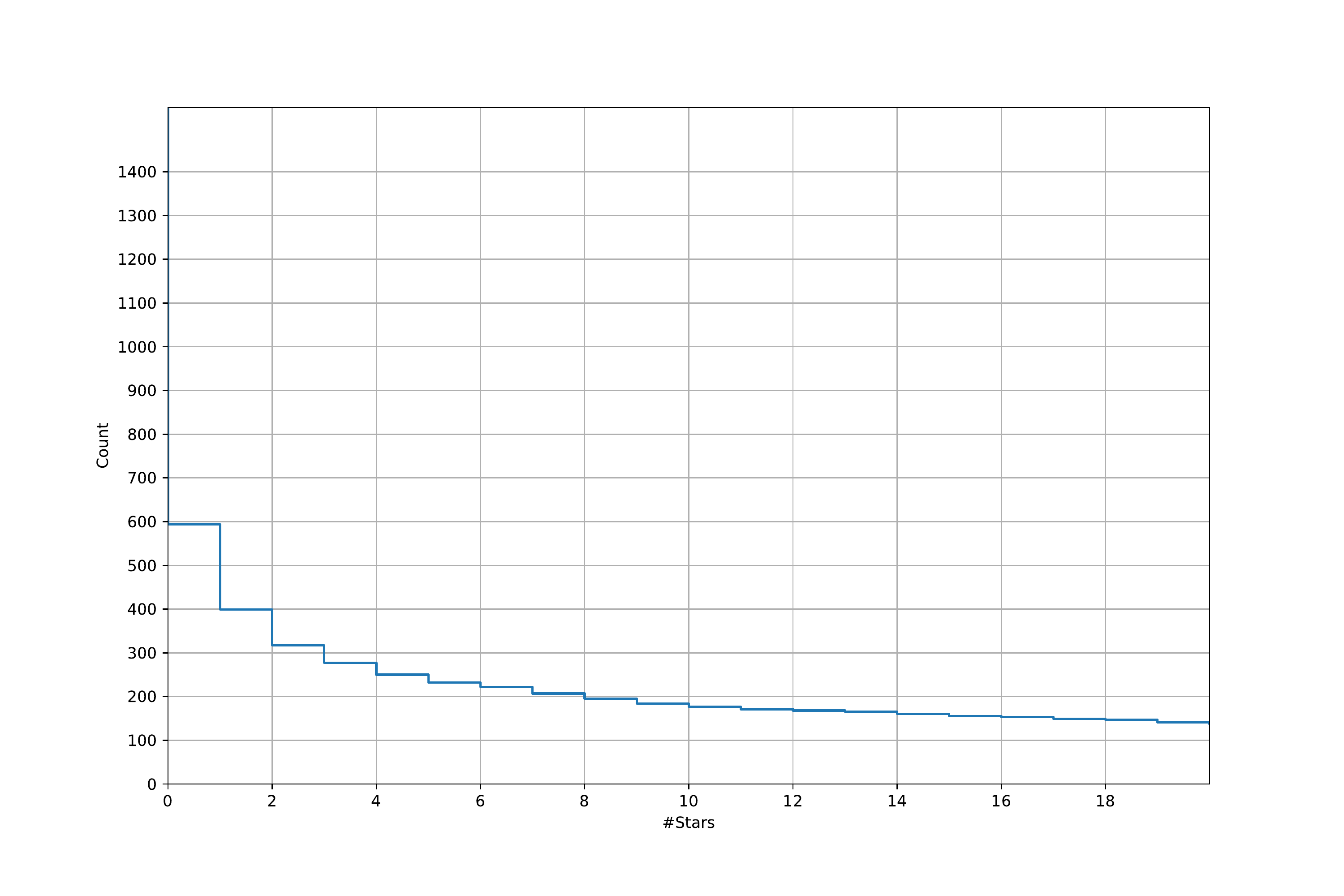}
\caption{Cumulative frequency curve base on \#Stars}
\label{img2}
\end{subfigure}
\caption{Cumulative frequency curve of \#Stars and \#Authors}
\label{img12}
\end{figure}
\end{appendices}

\end{document}